\documentclass[%
 reprint,
superscriptaddress,
showpacs,
 amsmath,amssymb,
 amsfonts,
 aps,
 pra,
showkeys
]{revtex4-1}

\usepackage{graphicx}
\usepackage{dcolumn}

\RequirePackage{atbegshi}
\usepackage{hyperref}

\usepackage{bbold}
\usepackage{verbatim}

\usepackage{subfigure}

\usepackage{units}
\usepackage[]{todonotes}

\usepackage[utf8]{inputenc}
\usepackage[T1]{fontenc}
\usepackage[ngerman,english]{babel}
\usepackage{slashed}
\usepackage{upgreek}
\usepackage{dsfont}

\usepackage{hyperref}
\hypersetup{
    colorlinks=false,
    citecolor=black,
    filecolor=black,
    linkcolor=black,
    urlcolor=black,
    linktocpage
}

\newcommand{\Tr}{\operatorname{Tr}}
\newcommand{\TrC}{\ensuremath{\Tr_c}}
\newcommand{\real}{\operatorname{Re}}

\newcommand{\ee}{\ensuremath{\textrm{e}}}
\newcommand{\ii}{\ensuremath{\textrm{i}}}

\newcommand{\NGroup}{\ensuremath{\text{N}}}
\newcommand{\NSigma}{\ensuremath{\text{N}_\sigma}}
\newcommand{\NTau}{\ensuremath{\text{N}_\tau}}

\newcommand{\Nc}{\ensuremath{\text{\NGroup}_\text{c}}}
\newcommand{\Nf}{\ensuremath{\text{\NGroup}_\text{f}}}

\newcommand{\VSpatial}{\ensuremath{V}}

\newcommand{\Loewe}{LOEWE-CSC}
\newcommand{\Lcsc}{L-CSC}

\newcommand{\clqcd}{CL\kern-.25em\textsuperscript{2}QCD}

\newcommand{\Ocl}{OpenCL}

\newcommand{\psibar}{\bar{\psi}}

\newcommand{\mpi}{\ensuremath{m_{\pi}}}
\newcommand{\mpiC}{\ensuremath{\mpi^c}}

\newcommand{\MFermion}{\ensuremath{D}}

\newcommand{\MFermionMinus}{\ensuremath{\MFermion^{-1}}}

\newcommand{\Action}{\ensuremath{\mathcal S}}

\newcommand{\SFermion}{\ensuremath{\Action_{\text{f}}}}

\newcommand{\SGauge}{\ensuremath{\Action_{\text{gauge}}}}

\newcommand{\G}{\ensuremath{g}}
\newcommand{\GSq}{\ensuremath{\G^2}}

\newcommand{\Order}{\ensuremath{\mathcal{O}}}

\newcommand{\Link}{\ensuremath{U}}

\newcommand{\LatSpacing}{\ensuremath{a}}
\newcommand{\LatMass}{\ensuremath{m}}

\newcommand{\Plaq}{\ensuremath{P_{\mu\nu}}}

\newcommand{\LatX}{\ensuremath{n}}

\newcommand{\LatY}{\ensuremath{m}}

\newcommand{\LatCoupling}{\ensuremath{\beta}}
\newcommand{\LatCouplingC}{\ensuremath{\LatCoupling_c}}

\newcommand{\Binder}{\ensuremath{B_4}}
\newcommand{\Skew}{\ensuremath{S}}
\newcommand{\Temp}{\ensuremath{T}}

\newcommand{\Mu}{\ensuremath{\mu}}
\newcommand{\MuI}{\ensuremath{\mu_i}}
\newcommand{\MuIc}{\ensuremath{\MuI^c}}

\newcommand{\GenObs}{\ensuremath{X}}

\begin{document}

\title{The $\Nf=2$ QCD chiral phase transition with Wilson fermions at zero and imaginary chemical potential}

\author{Owe Philipsen}
 \email{philipsen@th.physik.uni-frankfurt.de}
\affiliation{
 Institut f\"{u}r Theoretische Physik - Johann Wolfgang Goethe-Universit\"{a}t, Germany\\
 Max-von-Laue-Str.~1, 60438 Frankfurt am Main
}
\affiliation{
John von Neumann Institute for Computing (NIC)
GSI, Planckstr.~1, 64291 Darmstadt, Germany
}
\author{Christopher Pinke}
 \email{pinke@th.physik.uni-frankfurt.de}
 \affiliation{
  Institut f\"{u}r Theoretische Physik - Johann Wolfgang Goethe-Universit\"{a}t, Germany\\
  Max-von-Laue-Str.~1, 60438 Frankfurt am Main
 }

\date{\today}

\begin{abstract}
The order of the thermal phase transition in the chiral limit of Quantum Chromodynamics (QCD) with two 
dynamical flavors of quarks is a long-standing issue and still not  known in the continuum limit. 
Whether the transition is first or second order has important implications for the QCD phase diagram 
and the existence of a critical endpoint at finite densities. 
We follow a recently proposed approach to explicitly determine the region of first order chiral transitions
at imaginary chemical potential, where it is large enough to be simulated, 
and extrapolate it to zero chemical potential with known critical exponents.
Using unimproved Wilson fermions on coarse $N_t=4$ lattices, the first order region turns out to be so large
that no extrapolation is necessary. The critical pion mass $m_\pi^c\approx 560$ MeV is by nearly a factor 10 larger
than the corresponding one using staggered fermions. Our results are in line with investigations of three-flavour
QCD using improved Wilson fermions and indicate that the systematic error on the two-flavour chiral
transition is still of order 100\%. 
\end{abstract}

\maketitle

\section{Introduction}
\label{ch:introduction}

Mapping out the phase diagram of Quantum Chromodynamics (QCD) as a function of temperature $T$
and baryon chemical potential $\mu_B$ is one of the most challenging 
tasks of modern particle physics.
As the strong interactions are inherently nonperturbative on hadronic energy scales, Lattice QCD (LQCD) 
is the only first principle approach to date, for which all systematic errors can  
eventually be removed.

\begin{figure*}
		\centering
		\begin{minipage}{.45\linewidth}
					\includegraphics[scale=1]{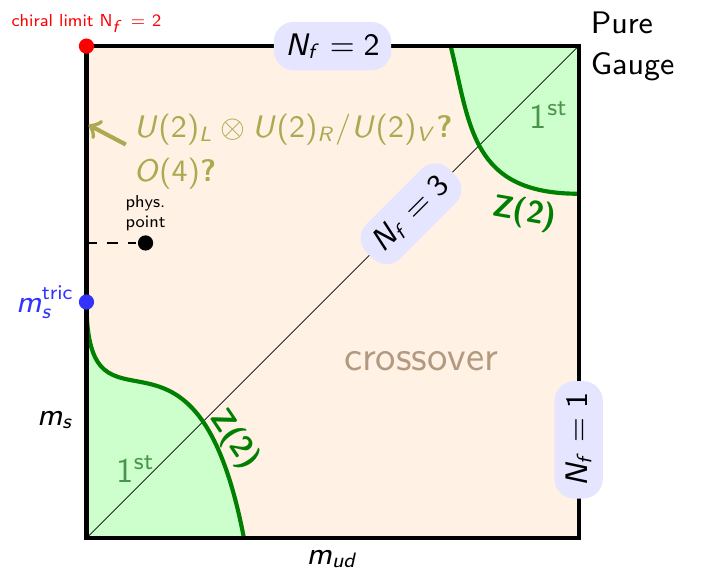}
		\end{minipage}
		~~
		\begin{minipage}{.45\linewidth}
					\includegraphics[scale=1]{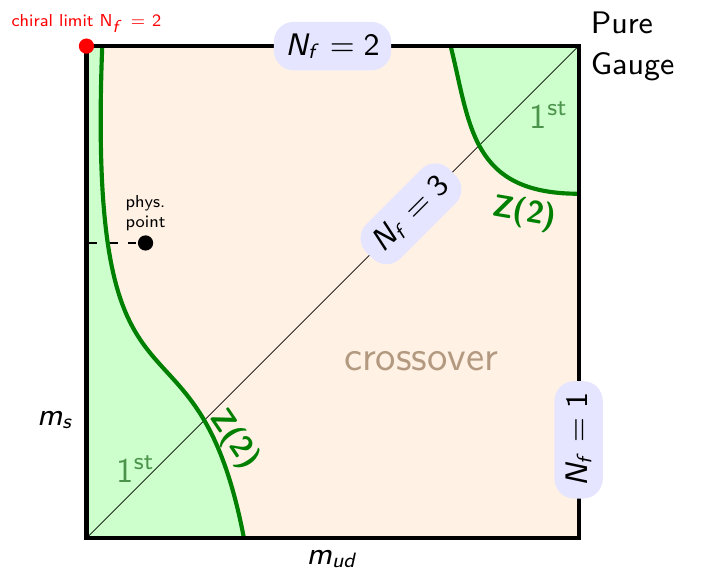}
		\end{minipage}
		\caption{Possible scenarios for the QCD phase transition at $\Mu=0$ as function of quark mass. See text for details.}
		\label{fig:columbiaPlot}
\end{figure*}

The order of  the thermal transition from a hadron gas to a 
quark gluon plasma changes as a function of the quark masses.
The qualitative situation at zero baryon chemical potential is depicted in Fig.~\ref{fig:columbiaPlot}.
Regions of first order phase transitions are seen on coarse lattices for three degenerate 
flavors of quarks ($\Nf=3$) with 
large and small masses.
These are center-symmetry breaking (deconfinement) and chiral symmetry restoring 
phase transitions, respectively.
At intermediate masses, including the physical point, the thermal transition proceeds by an analytic crossover.
The first order and crossover regions are separated by second order lines in the 
3D Ising universality class ($Z(2)$),
which have 
been mapped out on coarse lattices (\cite{deForcrand:2006pv,Fromm:2011qi,Saito:2011fs} and 
references therein).
However for two flavors of quarks, the nature of the chiral phase transition (upper left corner 
of Fig.~\ref{fig:columbiaPlot}) is particularly difficult to clarify
because chiral fermions cannot be simulated easily.
In the massless, chiral limit, the transition may be of either first or second order \cite{Pisarski:1983ms,Butti:2003nu}, 
corresponding to the two different scenarios in Fig.~\ref{fig:columbiaPlot}.
Which option is actually realized in QCD 
is a long-standing and controversial issue, for a recent overview see e.g.~Ref.~\onlinecite{Meyer:2015}.
Settling this issue is important, since the nature of the chiral transition at zero density also 
has implications for the physical QCD phase diagram
at finite baryon density, which cannot be simulated 
directly because of the sign-problem of LQCD. 
In particular, it influences the possibility of a critical endpoint at moderate densities \cite{Rajagopal:2000wf}.

A standard approach to address this question is to simulate the $\Nf=2$ crossover region at 
successively decreasing pion masses and search for 
scaling behavior related to a critical point. However, such studies are computationally expensive
and often inconclusive, because of the similarity of the critical exponents distinguishing the second order
points of the two
scenarios. For a recent discussion, see e.g.~Refs.~\onlinecite{Meyer:2015,Bonati:2014kpa}.
Recently, an alternative approach was used in Ref.~\onlinecite{Bonati:2014kpa}, which employs the fact that
the first order chiral transition region widens when an imaginary chemical potential is switched on,
and thus can be simulated directly. The second order boundary between the crossover and first order 
region can then be extrapolated to zero density with known exponents, which are
induced by the Roberge-Weiss symmetry. 
Using unimproved staggered fermions on coarse $\NTau=4$ lattices, it was indeed established that 
the transition is of first order in the chiral limit. (A similar strategy is followed using additional heavy flavours
in \cite{Ejiri:2015vip}, though no tricritical scaling has been reported there as yet).

It must be stressed that so far there is no continuum extrapolation for any of these features. On the contrary, 
it is becoming clear that the locations of the critical boundary lines display particularly strong cut-off effects.
In particular, the first order chiral transition region for staggered fermions 
shrinks drastically on finer $\NTau=6$ lattices \cite{deForcrand:2007rq,Endrodi:2007gc}, and can only be 
bounded when using improved staggered fermions \cite{Ding:2015pmg}. 
By contrast and indicating the size of cut-off effects, the chiral first order region is found to stay rather wide
when using improved Wilson fermions \cite{Takeda:2014nta}. The purpose of this paper is to investigate 
cut-off effects on the $\Nf=2$ chiral transition region by repeating the study of Ref.~\onlinecite{Bonati:2014kpa} with 
unimproved Wilson fermions, starting from previous studies at imaginary 
chemical potential in Refs.~\onlinecite{Philipsen:2014rpa,Cuteri:2015qkq}. The reason we use unimproved
Wilson fermions is two-fold. On a conceptual level, one can be sure that there are no unphysical modifications
to the phase structure due to improvement terms. On a practical level, if we wish to quantify cut-off effects 
and eventually remove them by extrapolation, it is necessary to see and control the chiral phase transition, rather
than just bounding it. 

We summarize QCD at imaginary chemical potential in Sec. \ref{ch:theory} and give technical details of our simulation setup in Sec.~\ref{ch:simulationDetails}.
Numerical results are presented in Sec.~\ref{ch:results}, followed by a discussion in Sec.~\ref{ch:discussion}.
\begin{figure*}
		\centering
		\begin{minipage}{.45\linewidth}
					\includegraphics[scale=.7]{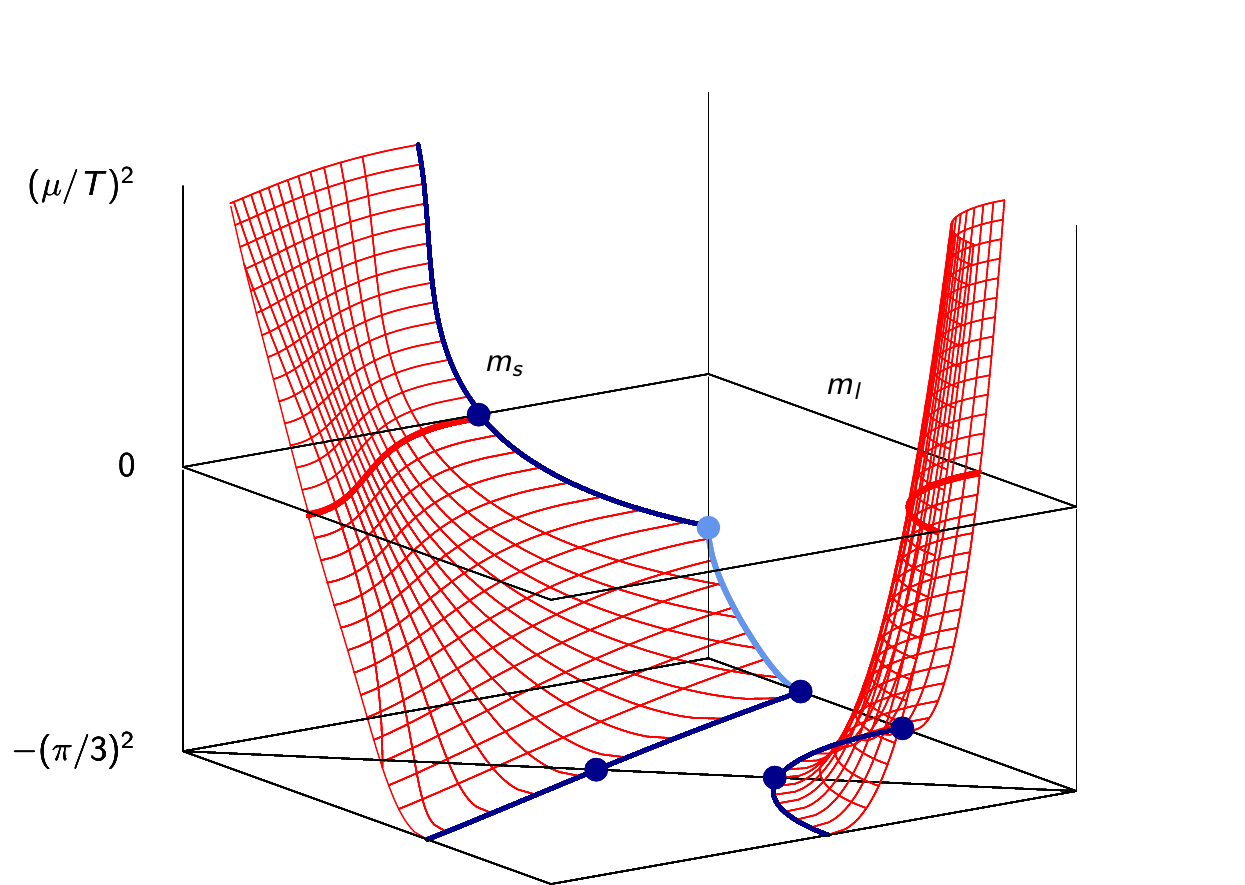}
		\end{minipage}
		~~
		\begin{minipage}{.45\linewidth}
					\includegraphics[scale=1]{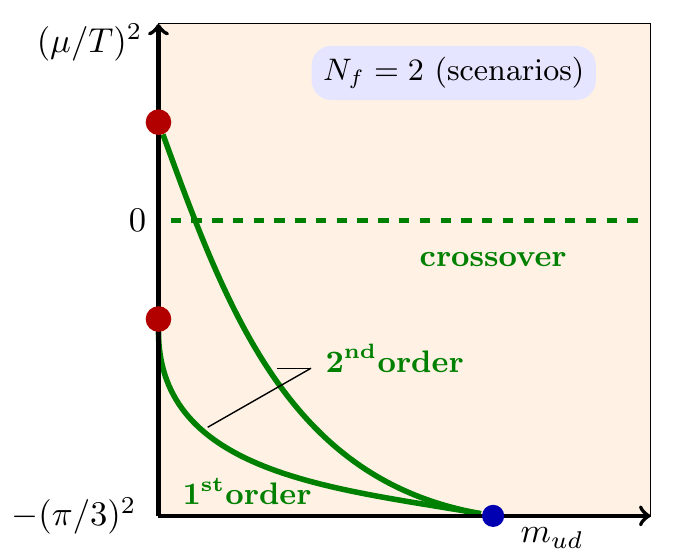}
		\end{minipage}
		\caption[]{Left: QCD phase diagram as function of $(\mu/T)^2$ as extension
of Fig.~1 (left). The red
surfaces mark second order transitions. Bold dark blue lines on the surfaces $\mu/T=i\pi/3$
and $m_{ud}=0$ are tricritical lines. The bold bright blue line on the
surface $m_s=\infty$ is the second order line we study.   
Right: The $N_f=2$ backplane ($m_s=\infty$). Solid dots depict
tricritical points and the two green lines describe scenarios I and II for the second order
transition line. Both figures follow Ref.~\onlinecite{Bonati:2014kpa}.
}
		\label{fig:columbiaPlot3dAndNf2Backplane}
\end{figure*}

\section{QCD at imaginary chemical potential}
\label{ch:theory}

At imaginary quark chemical potential $\mu=\ii \MuI$  ($\mu = \mu_B/3$) 
the sign problem is absent and standard simulation algorithms can be applied.
QCD possesses a rich phase structure in this region, 
which depends on the number of flavors \Nf\ and the quark mass $m$.
The partition function is an even function of \Mu\ due to $CP$-symmetry, and it is periodic in $\mu/T$ with 
period $2\pi/\Nc$  due to gauge symmetry and the anti-periodic boundary conditions of fermions
in the temporal direction \cite{Roberge:1986mm}.
As a consequence, critical values $\MuIc/T = (2k+1)\pi/\Nc\ (k \in \mathbb N)$ mark the boundaries 
between adjacent, physically equivalent $Z(\Nc)$ center sectors of the gauge group (throughout the paper we 
use $\Nc = 3$).
The transitions in the $\mu_i$-direction between these sectors are called Roberge-Weiss (RW) transitions.
For low temperatures, the RW transition is a smooth crossover, whereas it becomes a first order transition for high \Temp\ \cite{Roberge:1986mm,deForcrand:2002ci,D'Elia:2002gd}.
Consequently, there is a  so-called RW endpoint, where these two distinct behaviors meet.
For small and large quark masses, the analytic continuation of the chiral and deconfinement transitions
also join this point, which then becomes a triple point. For intermediate masses, where there is no chiral or
deconfinement transition, it is instead a second order endpoint. 
Hence, the nature of the RW endpoint depends on the masses and the number of flavors just as the order 
of the transition at $\mu=0$ does. 

This is the content of Fig.~\ref{fig:columbiaPlot3dAndNf2Backplane} (left), which represents 
Fig.~\ref{fig:columbiaPlot} (left) enlarged by an additional $\mu^2$-axis. 
The value $\mu_i/T=\pi/3$ denotes the RW-plane with its regions of triple point behavior and
second order endpoint behavior, separated by tricritical lines.
The critical lines at $\mu=0$ bounding the chiral and deconfinement transitions
continue as critical surfaces to imaginary chemical potential and terminate in these 
tricritical lines  \cite{deForcrand:2010he}.
This phase structure has been mapped out in recent years, and is qualitatively the same using staggered \cite{deForcrand:2002ci,D'Elia:2002gd,deForcrand:2003vyj,deForcrand:2006pv,deForcrand:2008vr,D'Elia:2009qz,deForcrand:2010he,Bonati:2010gi,Bonati:2014kpa,Bonati:2016pwz} or Wilson fermions \cite{Nagata:2011yf, Alexandru:2013uaa, Wu:2013bfa, Philipsen:2014rpa, Cuteri:2015qkq}. 

Our interest now is in the $\Nf=2$ backplane, shown in 
Fig.~\ref{fig:columbiaPlot3dAndNf2Backplane} (right).
More specifically, leaving the critical $\MuI$-value of the RW-transition (bottom of the figure), 
a line of second order transitions departs from the tricritical point, 
separating regions of first order transitions from crossover regions.
This line has to terminate in another tricritical point at $m_{ud}=0$. 
In the vicinity of tricritical points, the functional form of the line is governed by tricritical scaling laws, 
which allows for its extrapolation to the chiral limit \cite{Bonati:2014kpa}.
There are two possible scenarios as shown in Fig. \ref{fig:columbiaPlot3dAndNf2Backplane} (right).
If the tricritical point at $m_{ud}=0$ is at negative values of $\Mu^2$, the chiral phase transition is second order.
On the other hand, if it is at positive values, there exists a first order region at $\Mu=0$ and 
the transition in the chiral limit must be first order, too.
In this way one can clarify the order of the chiral limit at zero chemical potential by mapping out the second 
order line. For staggered fermions on $\NTau=4$ lattices, it was found to be of first order \cite{Bonati:2014kpa}. 
In this work we apply the same strategy using Wilson fermions.

\section{Simulation details}
\label{ch:simulationDetails}

We employ the same numerical setup as for our previous studies of Wilson fermions at imaginary 
chemical potential, described in Refs.~\cite{Philipsen:2014rpa} and \cite{Cuteri:2015qkq}.
In the gauge sector we use the standard Wilson gauge action,
\begin{equation}
	\label{WilsonGaugeAction}	
	\SGauge = \LatCoupling \sum_\LatX  \sum_{\mu, \nu > \mu} \left\{ 1 - \real \TrC (\Plaq(\LatX) )\right\} \;,\nonumber
\end{equation}
with plaquette \Plaq\ and lattice coupling $\LatCoupling = \frac{2 \Nc}{\GSq}$,
lattice sites \LatX, Lorentz indices $\mu, \nu$ and bare coupling $\GSq$.
In the fermionic sector we consider two flavors of mass-degenerate quarks with the standard Wilson action
\begin{equation}
	\label{SFermLatGen}
	\SFermion [\psibar, \psi, \Link ] = \LatSpacing^{4} \sum_{\Nf}\sum_{\LatX,\LatY} \psibar(\LatX)\ \MFermion(\LatX, \LatY) [\Link ]\ \psi(\LatY)\;, \nonumber
\end{equation}
and fermion matrix
\begin{align}
	\label{MWilson}
	\MFermion(\LatX, \LatY) 	=&\  \delta_{\LatX\LatY}
	- \kappa \sum_{i} (1 - \gamma_i)\ee^{\LatSpacing\Mu\delta_{|i|,0}\text{sgn}(i)}\Link_{\pm i}(\LatX)\delta_{\LatX+\hat{i},\LatY}\;.\nonumber
\end{align}
Here, the shorthand notation $\gamma_{-\mu} = - \gamma_\mu$ and $\Link_{-\mu}(\LatX) = \Link^{\dagger}_\mu(\LatX - \vec\mu)$ has been used.
The bare fermion mass \LatMass\ sets the value of the hopping parameter
\begin{equation}
	\label{LatMassWilson}
	\kappa = (2(\LatSpacing \LatMass + 4))^{-1} \;.\nonumber
\end{equation}
Finite temperature on the lattice is given by 
\begin{equation}
	\label{LatTemp}
	\Temp = 1/\left( \LatSpacing(\LatCoupling)\NTau \right)\;.\nonumber
\end{equation}

All our numerical simulations have been performed using the \emph{publicly available}~\cite{CL2QCD} \Ocl~\cite{opencl} based code \clqcd~\cite{Bach:2012iw,Philipsen:2014mra}, which is optimized to run efficiently on GPUs. In particular, the \Loewe~\cite{Bach2011a} at Goethe-University Frankfurt and the \Lcsc~\cite{L-CSC} at GSI in Darmstadt have been used.

We work at fixed temporal lattice extent $\NTau = 4$, leaving the RW-plane $\MuIc=\pi\Temp/3$ investigated in Ref. \onlinecite{Philipsen:2014rpa} at the same \NTau.
We work at four different values of the bare quark mass, parametrized by $\kappa = 0.165, 0.17, 0.175$ and 
$0.18$ to account for the shift of the critical line towards smaller masses.
In order to locate the critical chemical potential for each bare quark mass, we scan in $a\Mu_i$ for each mass.
We also performed a mass scan at $\Mu=0$ in $\kappa = 0.175, 0.1775, 0.18$ and $0.1825$. 

For all parameter sets, temperature scans were carried out on $\NSigma = 12$ lattices, locating the (pseudo-)critical 
gauge coupling \LatCouplingC\ with a $\Delta\LatCoupling$ of at most $0.001$ around \LatCouplingC.
Simulations on larger volumes $\NSigma = 16$ and $20$ were added for finite size scaling, corresponding 
to aspect ratios $\NSigma/\NTau$ of 3, 4 and 5, respectively.
In order to accumulate statistics, we simulated four independent Monte Carlo 
chains for each parameter set, with acceptance rates of the order of $75\%$ for each run.
The autocorrelation on the data was estimated using a python implementation \footnote{See github.com/dhesse/py-uwerr.} of the Wolff method \cite{Wolff:2003sm}.
After discarding 5k to 10k trajectories for thermalization, 40k to 100k trajectories were collected for each individual Monte-Carlo chain, such that there are $\mathcal{O}(100)$ statistically 
independent configurations in the critical region.
Observables were measured after each trajectory. 
Additional \LatCoupling-points have been generated 
using Ferrenberg-Swendsen reweighting \cite{Ferrenberg:1989ui}.
For scale-setting purposes, $\Temp=0$ simulations at or close to certain critical parameters have been performed.
The scale itself is then set by the Wilson flow parameter $w_0$ using the publicly available code described in Ref.~\onlinecite{Borsanyi:2012zs}.
This method is very efficient and fast.
In addition, the pion mass \mpi\ was determined along the critical line using these vacuum configurations.

\begin{table*}
\caption{Results for the critical coupling \LatCouplingC\ and corresponding \Binder\ value. 
\LatCouplingC\ has been determined from the data by a vanishing skewness.
For \LatCouplingC, a constant, conservative error of $0.0005$ has been assigned given our resolution in the simulated points of $\Delta\beta=0.001$.
Values for $\MuI/\Temp$ are chosen such that the simulation points gradually leave the RW-plane  (see text) and follow the $Z(2)$-line to smaller $\MuI/\Temp$ values as the (bare) mass is lowered.
In addition, simulations at $\MuI/\Temp=0$ have been added.\\
}
\label{tab:simulations}
\begin{tabular}{|c|c|c|c|c|c|c|c|}
\hline
$\kappa$ & $\MuI/\Temp$ & \LatCouplingC(\NSigma =12) & $\Binder(\LatCouplingC, \NSigma =12)$
   & $\LatCouplingC(\NSigma =16)$ & $\Binder(\LatCouplingC, \NSigma =16)$& $\LatCouplingC(\NSigma =20)$ & $\Binder(\LatCouplingC,\NSigma =20$)\\
\hline
0.165  & 0.984     &  5.2439(5) & 1.564(19)  & 5.2440(5) & 1.492(20)  &  5.2439(5) & 1.458(30)\\
       & 0.890   &  5.2356(5) & 1.878(48)  & 5.2354(5) & 1.874(52)  &  5.2356(5) & 2.062(62) \\
       & 0.796    &  5.2287(5) & 2.085(55)  & 5.2284(5) & 2.265(59)  &  5.2283(5) & 2.629(79) \\
       & 0.733    &  5.2241(5) & 2.246(108) & 5.2243(5) & 2.068(111) & 5.2245(5) & 2.861(298) \\
       & 0.576    &  5.2159(5) & 2.391(145) & 5.2159(5) & 2.505(151) & - & - \\
       & 0.419    &  5.2094(5) & 2.543(115) & 5.2096(5) & 2.897(128) & - & -  \\
\hline
 0.17  & 0.890     &  5.1561(5) & 1.417(23)  & 5.1560(5) & 1.269(108) & 5.1561(5) & 1.184(17) \\
       & 0.733    &  5.1459(5) & 1.728(32)  & 5.1459(5) & 1.819(38)  & 5.1459(5) & 1.847(116) \\
       & 0.576   &  5.1383(5) & 1.926(38)  & 5.1385(5) & 2.118(35)  & 5.1385(5) & 2.340(48)  \\
\hline
0.175  & 0.733   &  5.0601(5) & 1.290(36) & 5.0596(5) & 1.215(108) & 5.0604(5) & 1.113(94) \\
       & 0.576  &  5.0533(5) & 1.522(33) & 5.0531(5) & 1.439(31)  & 5.0531(5) & 1.459(41) \\
       & 0.419  &  5.0481(5) & 1.774(36) & 5.0482(5) & 1.859(48)  & 5.0480(5) & 1.991(66) \\
       & 0          &  5.0426(5) & 2.035(32) & 5.0427(5) & 2.113(37)  & 5.0425(5) & 2.383(44) \\
\hline
0.1775 & 0          &  4.9981(5) & 1.888(34) & 4.9981(5) & 2.066(46) & 4.9981(5) & 1.864(31)  \\
\hline
0.18   & 0.419  &  4.9568(5) & 1.466(56) & 4.9568(5) & 1.357(41) & 4.9567(5) & 1.370(158) \\
       & 0.262  &  4.9537(5) & 1.560(44) & 4.9539(5) & 1.445(52) & 4.9538(5) & 1.492(34)  \\
       & 0.105   &  4.9523(5) & 1.704(30) & 4.9523(5) & 1.683(30) & 4.9522(5) & 1.683(35)  \\
       & 0         &  4.9521(5) & 1.714(26) & 4.9520(5) & 1.812(35) & 4.9519(5) & 1.863(49)  \\
\hline
0.1825 & 0          &  4.9045(5) & 1.604(30) & 4.9043(5) & 1.605(34) & 4.9044(5) & 1.401(19) \\
\hline
\end{tabular}
\end{table*}

\section{Results}
\label{ch:results}

\begin{figure}
 		\centering
   \includegraphics[scale=0.6]{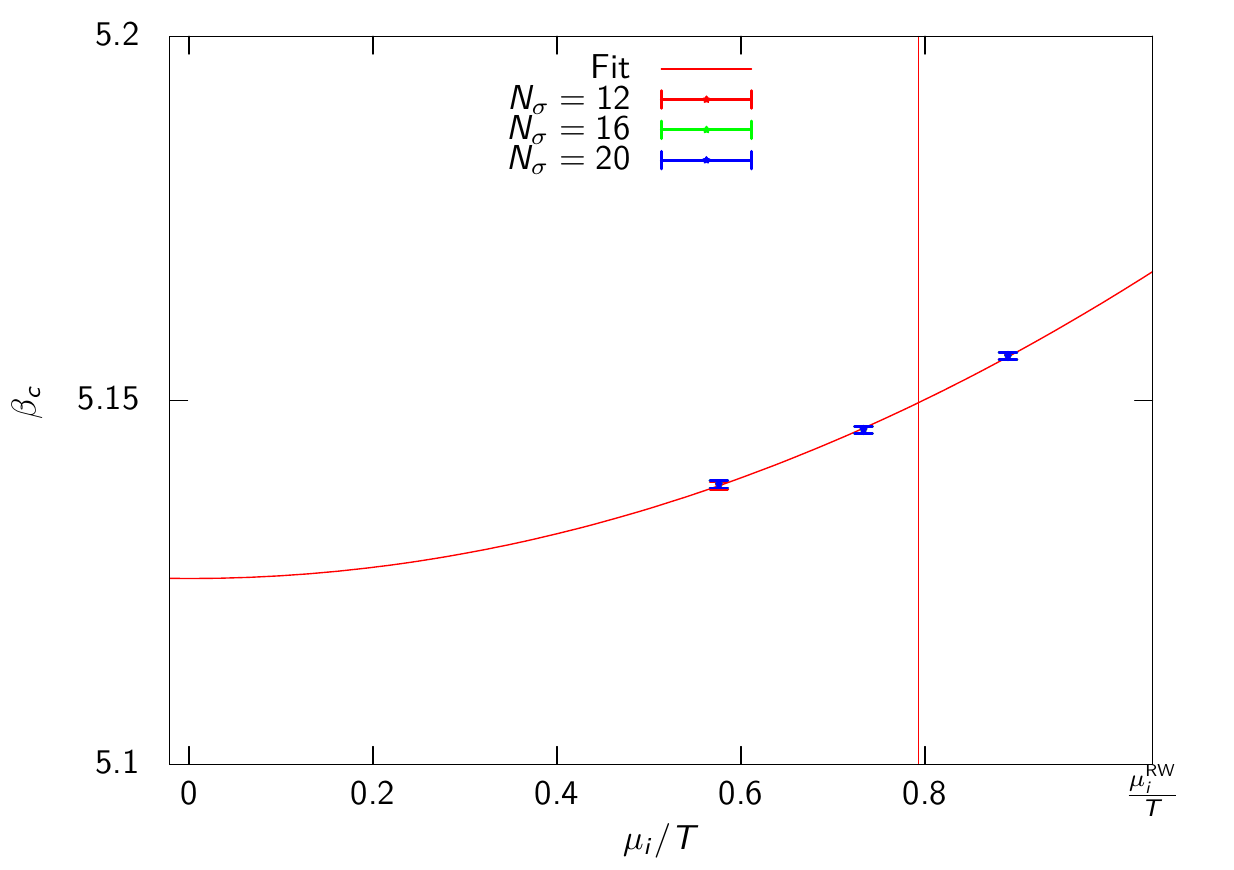}
 		\caption{Critical temperature as function of $\MuI/T$ for $\kappa=0.17$. The vertical line indicates the $Z(2)$-critical value of $\MuI/T$ (See Table~\ref{tab:fitResults1}).}
 		\label{fig:betac_fit}
\end{figure}

\begin{figure*}[t]
	\centering
	\begin{minipage}{.45\linewidth}
				\includegraphics[scale=0.6]{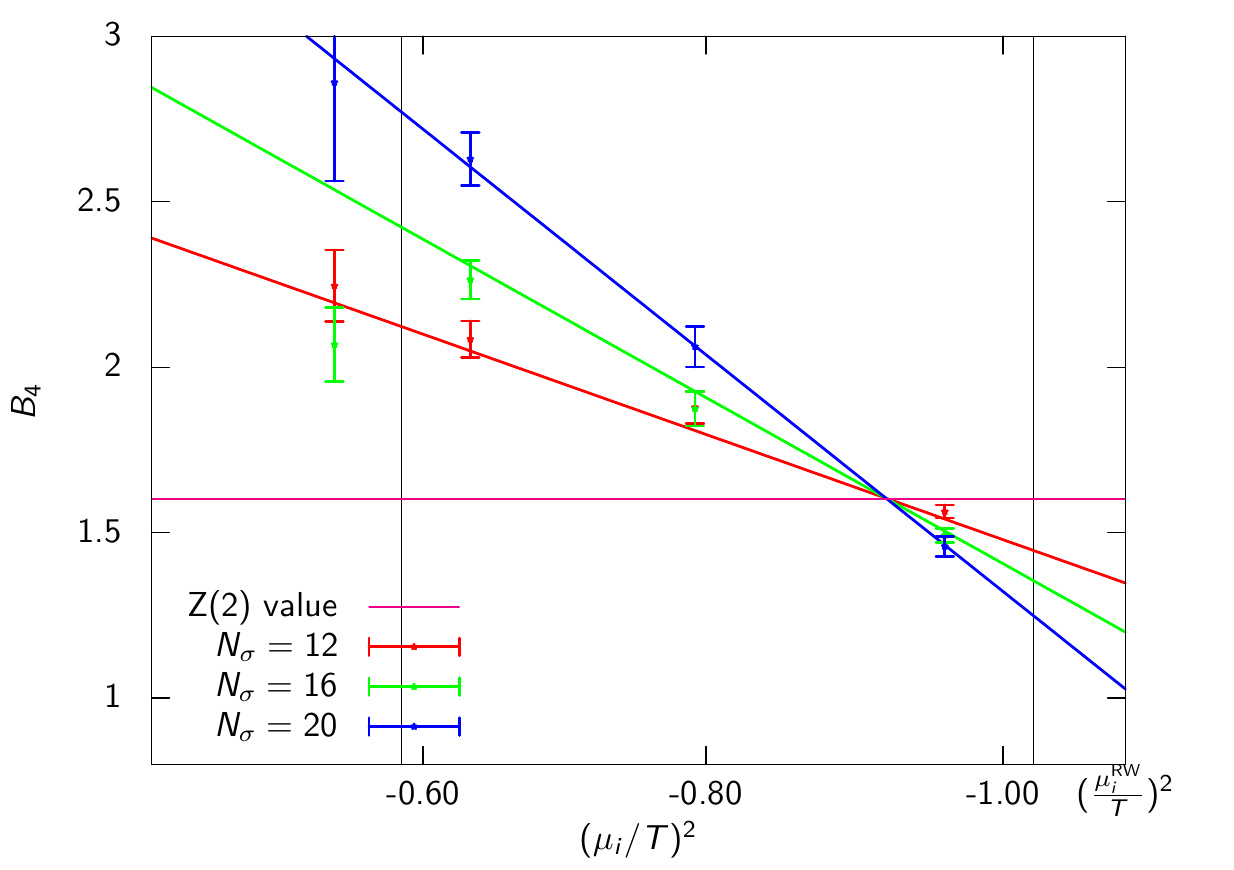}
	\end{minipage}~~
	\begin{minipage}{.45\linewidth}
				\includegraphics[scale=0.6]{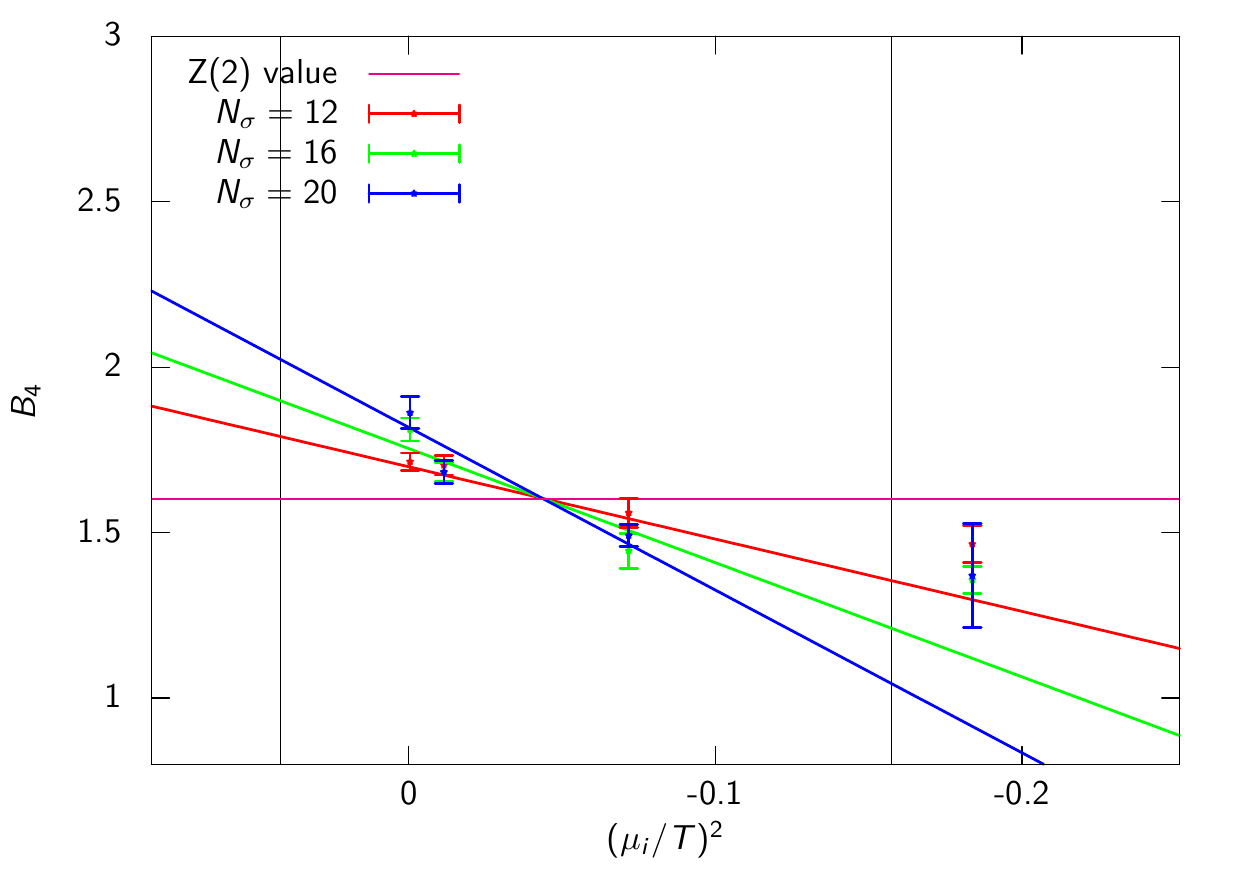}
	\end{minipage}
	\caption{Finite size scaling of \Binder\ and fits for $\kappa=0.165$ (left) and $\kappa=0.18$ (right). The vertical lines indicate the fit ranges.}
	\label{fig:results1}
\end{figure*}

We define a (pseudo-)critical coupling \LatCouplingC\ (corresponding to some temperature 
$\Temp_c$) by the vanishing of the skewness
\begin{equation}
	\Skew\ (\beta_c) =  \langle (\GenObs - \langle \GenObs \rangle)^{3}\rangle_{\beta_c} /  
	\langle	(\GenObs - \langle \GenObs \rangle )^{2}	\rangle_{\beta_c}^{3/2} =0 \nonumber
\end{equation}
of a suitable observable \GenObs.
In a study of the chiral transition it is natural to use the chiral condensate
\begin{equation}
	\label{pbp_1}
	X=\bar{\psi}\psi= \Nf\ \Tr{\MFermionMinus}\;. \nonumber
\end{equation}
As an example, we show $\beta_c(\mu_i/T)$ for $\kappa=0.17$ in Fig.~\ref{fig:betac_fit}. Note that, because
of the reflection symmetry of the partition function in $\mu$, the critical coupling is an even function
of chemical potential. For chemical potential values up to the RW-plane it is well fitted by 
the leading quadratic Taylor term (cf.~also \cite{deForcrand:2002ci,D'Elia:2002gd}).
This allows to interpolate between simulation points and zero chemical potential to obtain preliminary
estimates for $\beta_c$ which are then numerically tuned to the desired precision.
As expected, the results show a decreasing critical coupling, and thus temperature, as the chemical 
potential approaches zero. The same holds if the (bare) quark mass is lowered.
See Table \ref{tab:simulations} for details.

We use the Binder cumulant \cite{Binder:1981sa} of the chiral condensate,
\begin{equation}
	\Binder(\GenObs) = \langle (\GenObs - \langle \GenObs \rangle)^{4}\rangle /  \langle	(\GenObs - \langle \GenObs \rangle )^{2}	\rangle^{2} \;,\nonumber
	\label{bindercum}
\end{equation}
evaluated at the coupling of vanishing skewness, i.e.~on the phase boundary, in 
order to extract the order of the transition as a function of quark mass and chemical potential, 
$\Binder(\beta_c;m,\MuI/\Temp,V)$.
In the thermodynamic limit $\VSpatial \rightarrow \infty$, it takes the values 1 for a first order transition and 3 
for an analytic crossover, respectively.
For the case of a second order transition in the $3D$ Ising universality class it takes the value $1.604$ \cite{deForcrand:2006pv}.
Hence, a discontinuity exists when passing from a first order region to a crossover region via a second order 
point. On finite volumes, this discontinuous step function is smeared out to a smooth function.
On sufficiently large volumes and in the vicinity of a critical point $(\beta_c,\mu_i^c/\Temp~(m))$, 
the Binder cumulant can be expanded in a finite size scaling variable according to
\begin{eqnarray}
\label{BinderScaling}
	\Binder(m,\MuI/\Temp,\NSigma) &=& \Binder(m,\MuI^c/\Temp,\infty)\\
	&+& b_1(m) \left\{(\MuI/\Temp)^2 - (\MuIc/\Temp)^2\right\} \NSigma^{1/\nu} + \ldots  \nonumber \;,
\end{eqnarray}
i.e.~it approaches the step function with a characteristic critical exponent $\nu$.
For the 3D Ising universality class, one has $\Binder \approx 1.604$ and 
$\nu \approx 0.63$.

\begin{figure}
	\includegraphics[scale=0.6]{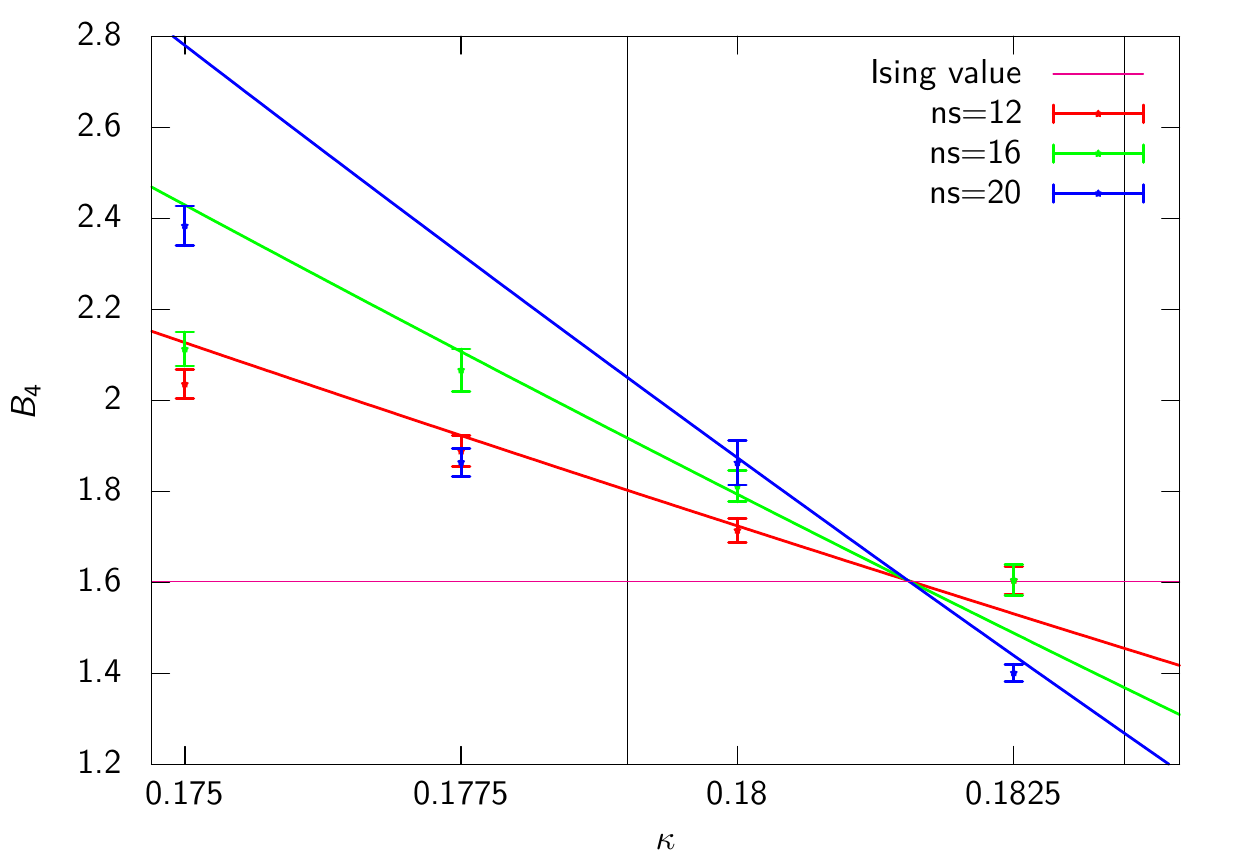}
	\caption{Finite size scaling of \Binder\ and fit for $\Mu=0$. The vertical lines indicate the fit ranges.}
	\label{fig:results3}
\end{figure}
Our simulated values for $\Binder(m,\mu_i^c/\Temp, \NSigma)$ are then fitted to this form, 
resulting in the fit parameters $b_1,\MuI^c/\Temp$, where
 $\MuIc/\Temp$ indicates the position of the critical point.
Examples are given in Fig.~\ref{fig:results1} and results for $\MuIc/\Temp$ are collected in Table \ref{tab:fitResults1}.
As can be seen in the figures, \Binder\ increases with volume if the transition is a crossover 
(left of the second order point), whereas it decreases in the first order region, 
ultimately approaching the infinite volume values 3 and 1, respectively.
As the (bare) quark mass is lowered, $\MuIc/\Temp$ decreases towards zero (see Table \ref{tab:fitResults1}).

Contrary to the situation with staggered quarks, the first order region in this case is wide enough to be simulated
everywhere and no extrapolation is necessary.
In order to locate the critical mass at $\mu=0$ directly, a scan in $\kappa$ was performed, with 
results presented in Table \ref{tab:fitResults2} and plotted in Fig.~\ref{fig:results3}.
Since here we  scan at fixed $\mu=0$ in $\kappa$, Eq. \eqref{BinderScaling} needs to be replaced by its analogue \cite{deForcrand:2003vyj}
\begin{eqnarray}
	\Binder(\kappa, \NSigma) &=& \Binder(\kappa,\infty) + k_1 \left\{\frac{1}{\kappa} - \frac{1}{\kappa_c}\right\} \NSigma^{1/\nu} + \ldots\;.
	\label{ScalingKappa}
\end{eqnarray}
Fitting the Binder data to Eq. \eqref{ScalingKappa} allows to extract $\kappa_c(\mu=0)$, 
see Table \ref{tab:fitResults2}.
Here we have interpolated the $\LatCouplingC$ data to get an estimate for  $\LatCouplingC(\kappa_c)$, 
similar to what is shown in Fig.~\ref{fig:betac_fit}.
Our results for the critical point in bare parameter space read:
\begin{eqnarray}
\kappa_c(\Mu=0) = 0.1815(1)\;,\;\LatCouplingC(\mu=0)=4.9228(1)\;.
\end{eqnarray}

\begin{table}
\caption{
Results for fits to Eq. \eqref{BinderScaling}. $\Binder(m,\MuI^c/\Temp,\infty)$ and $\nu$ have been set to $1.602$ and $0.6301$ throughout, respectively.\\
}
\begin{tabular}{|c|c|c|c|c|}
\hline
$\kappa$ & $\MuI^c/\Temp$ & $b_1$ & fit range & $\chi^2$  \\
\hline
0.165  & 0.9632(56) & -0.0293(13) & [$0.80:0.98$] & 0.85 \\
0.17   & 0.7924(24)  & -0.0219(6) & [$0.58:0.89$] & 0.42 \\
0.175  & 0.5292(60) & -0.0310(30) & [$0.42:0.58$] & 0.56 \\
0.18   & 0.2040(88) & -0.0443(48) & [$0.00:0.26$] & 1.94 \\
\hline
\end{tabular}
\label{tab:fitResults1}
\caption{
Results for fits at $\LatSpacing\Mu=0$ to Eq. \eqref{ScalingKappa}.
$\Binder(\kappa,\infty)$ and $\nu$ have been set to $1.602$ and $0.6301$, respectively.\\
}
\begin{tabular}{|c|c|c|c|}
\hline
$\kappa_c$ &  $b_1$ & fit range & $\chi^2$  \\
\hline
0.1815(1)  & 0.0492(39)   & [0.1800:0.1825] & 5.59 \\
\hline
\end{tabular}
\label{tab:fitResults2}
\end{table}

\begin{table*}[htbp]
\caption{
Overview of the $T=0$ simulations performed on $16^3 \times 32$ lattices.
	$w_0/a$ has been determined and converted to physical scales using the publicly available code described in Ref.\ ~\onlinecite{Borsanyi:2012zs}.
   For the pion mass determination, eight point sources per configuration have been used on $\mathcal{O}(400)$ uncorrelated configurations.
   The $\LatSpacing\mpi$  measurements for all $\kappa$ but $0.1575$ and $0.1815$ are taken from Ref. \onlinecite{thesisDepta}. 
	 The table also contains the lattice spacing and the pion mass in physical units and, in the last column, the temperature of the corresponding finite temperature ensemble with $\NTau=4$.\\ }
\begin{tabular}{|c|c||c|c||c|c||c|}
\hline
			$\kappa$ & $\LatCoupling$ & $w_0/a$     & $a\mpi$  & a[fm]    & $\mpi$ [MeV]& T [MeV] \\ \hline
			0.1815   & 4.9228         & 0.56418(9)  & 0.8828(3) & 0.311(3) & 560(6)      & 159(2)  \\ \hline			
			0.1800   & 4.9519         & 0.56738(5)  & 0.9076(2) & 0.309(3) & 579(6)      & 159(2)  \\ \hline
			0.1750   & 5.0519         & 0.58381(7)  & 0.9655(2) & 0.301(3) & 634(7)      & 164(2)  \\ \hline
			0.1700   & 5.1500         & 0.60973(10) & 1.0059(2) & 0.288(3) & 690(7)      & 171(2)  \\ \hline
			0.1650   & 5.2420         & 0.64801(16) & 1.0421(2) & 0.271(3) & 759(8)      & 182(2)  \\ \hline
			0.1575   & 5.3550         & 0.71045(26) & 1.1426(17) & 0.246(3) & 913(9)      & 200(2)  \\ \hline
\end{tabular}
\label{tab:scaleSetting_long}
\end{table*}

In order to compare with results from different discretizations or finer lattices in the future, we need
to convert the critical line to physical units.
The results of the scale setting procedure are summarized in Table \ref{tab:scaleSetting_long}.
(Also included are the re-evaluated results for $\kappa=0.1575$ from Ref.~\onlinecite{Philipsen:2014rpa}, 
which were originally carried out using a different scale setting method.)
The results show that, in terms of pion masses, the first order region is very large, with the 
smallest critical pion mass at zero density being $\mpiC(\Mu=0) \approx 560 \text{ MeV}$.
This result differs from an earlier one with unimproved Wilson fermions which was, however,  based on a different definition of the critical point \cite{Iwasaki:1996zt}. At this point, the critical boundary line does not yet fall
into the scaling region of the upper tricritical point and no controlled extrapolation to positive $\mu^2$
is possible.

Note that for simulations at fixed $\NTau$ the lattices coarsen going to lower masses, 
since \LatCouplingC\ decreases.
However, all lattices considered in this work are very coarse, with $\LatSpacing \gtrsim 0.25$ fm.
Because of this, large discretization artifacts are to be expected. On the other hand, our volumes satisfy
$\mpi L > 5$ for all our parameter sets, so that finite size effects are negligible.

\section{Discussion}
\label{ch:discussion}

\begin{figure}[b]
 		\centering
   \includegraphics[scale=0.7]{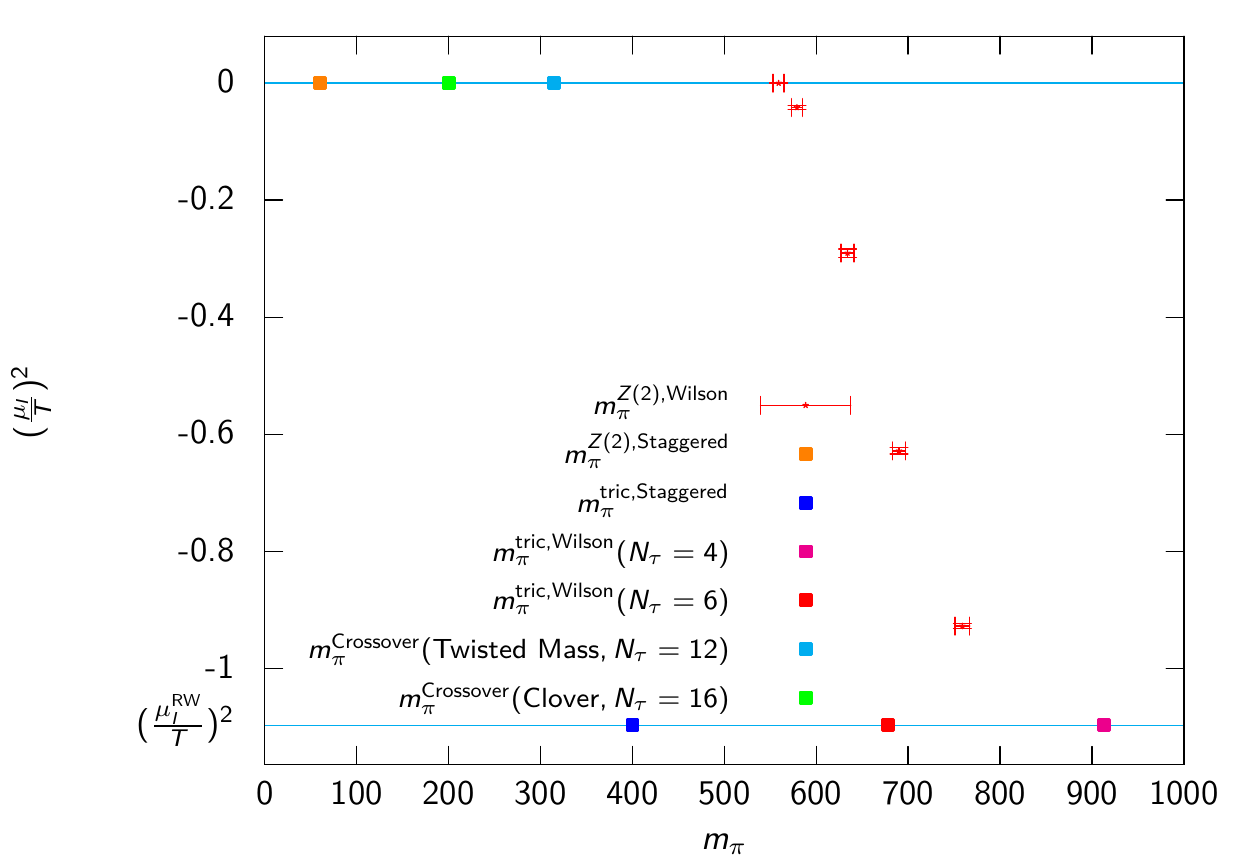}
 		\caption{$Z(2)$ line in the \mpi-$\Mu^2$ plane. See text for details.}
 		\label{fig:mpiVsMuSq}
\end{figure}

Our findings are summarized in Fig.~\ref{fig:mpiVsMuSq}, which shows the critical $Z(2)$ line separating
the regions of first order chiral transitions from analytic crossovers for $\Nf=2$ QCD on $\NTau=4$ lattices
in the \mpi-$(\Mu/\Temp)^2$ plane. As expected from the staggered results \cite{Bonati:2014kpa},
leaving the RW-plane towards $\mu=0$ the first order region shrinks as $(\mu/\Temp)^2$ grows. 
However, the lowering of \mpiC\ is relatively mild and we end up with a large region 
of first order chiral phase transitions at zero density, corresponding to scenario II
in Fig.~\ref{fig:columbiaPlot3dAndNf2Backplane} (right).
For comparison, the $\mu=0$ and $\mu_i=\pi T/3$ points of the critical line for
staggered fermions \cite{Bonati:2014kpa} are also shown in Fig.~\ref{fig:mpiVsMuSq}.
While the qualitative behavior is thus the same for both discretizations, the first order region is much wider 
for Wilson fermions. Assuming that both discretization schemes are fundamentally sound and lead to the
same continuum limit, we must conclude that the cut-off effect on the chiral critical pion mass at a lattice spacing
of $a\approx 0.25$ fm is of the order of $100\%$.

To place our results into context, the figure also shows two simulation points used in studies with 
$\Order(\LatSpacing)$-improved Wilson fermions on much finer lattices with
$\NTau = 12$ \cite{Burger:2011zc} and $16$ 
\cite{Brandt:2013mba}, where the thermal transition has been identified to be an analytic crossover.
These points may thus be taken as an upper bound for the critical pion mass in Wilson-type discretizations,
i.e.~the wide first order region is to a large extent due to discretization effects.
Indeed, in a recent study of the RW endpoint on $\NTau=6$ lattices \cite{Cuteri:2015qkq}, 
it was shown that the tricritical point in the RW-plane moves to lower masses by around 70\% with the unimproved action, as also shown in Fig.~\ref{fig:mpiVsMuSq}.
Assuming a similar shift at $\mu=0$ would put $\mpiC(\Mu=0, \NTau=6)$ to $\sim 400$ MeV and thus
in the vicinity to those crossover points.
Our findings are in qualitative accord with other investigations.
A recent study with $\Order(\LatSpacing)$-improved Wilson-Clover fermions determined a similarly 
large \mpiC\ of around 880 MeV for $\Nf=3$ on $\NTau=4$ lattices \cite{Jin:2014hea}.
Taken the improved and unimproved results together,
this suggests that the $\Order(\LatSpacing)$ effects are far from dominant on $\NTau=4$ lattices.

Altogether this suggests a very small or even vanishing $\mpiC(\Mu=0)$ in the continuum limit.
Being able to explicitly simulate the critical boundary of the transition region with the help 
of imaginary chemical potential and studying its change with the lattice spacing might in the future
allow for a continuum extrapolation.

\acknowledgments

O. P. and C. P. are supported by the Helmholtz  International Center for FAIR within the LOEWE program of the State of Hesse.
C.P. thanks the staff of \Loewe\ and \Lcsc\ for their support.

\bibliographystyle{apsrev4-1}
\bibliography{./literature.bib}

\begin{thebibliography}{43}%
\makeatletter
\providecommand \@ifxundefined [1]{%
 \@ifx{#1\undefined}
}%
\providecommand \@ifnum [1]{%
 \ifnum #1\expandafter \@firstoftwo
 \else \expandafter \@secondoftwo
 \fi
}%
\providecommand \@ifx [1]{%
 \ifx #1\expandafter \@firstoftwo
 \else \expandafter \@secondoftwo
 \fi
}%
\providecommand \natexlab [1]{#1}%
\providecommand \enquote  [1]{``#1''}%
\providecommand \bibnamefont  [1]{#1}%
\providecommand \bibfnamefont [1]{#1}%
\providecommand \citenamefont [1]{#1}%
\providecommand \href@noop [0]{\@secondoftwo}%
\providecommand \href [0]{\begingroup \@sanitize@url \@href}%
\providecommand \@href[1]{\@@startlink{#1}\@@href}%
\providecommand \@@href[1]{\endgroup#1\@@endlink}%
\providecommand \@sanitize@url [0]{\catcode `\\12\catcode `\$12\catcode
  `\&12\catcode `\#12\catcode `\^12\catcode `\_12\catcode `\%12\relax}%
\providecommand \@@startlink[1]{}%
\providecommand \@@endlink[0]{}%
\providecommand \url  [0]{\begingroup\@sanitize@url \@url }%
\providecommand \@url [1]{\endgroup\@href {#1}{\urlprefix }}%
\providecommand \urlprefix  [0]{URL }%
\providecommand \Eprint [0]{\href }%
\providecommand \doibase [0]{http://dx.doi.org/}%
\providecommand \selectlanguage [0]{\@gobble}%
\providecommand \bibinfo  [0]{\@secondoftwo}%
\providecommand \bibfield  [0]{\@secondoftwo}%
\providecommand \translation [1]{[#1]}%
\providecommand \BibitemOpen [0]{}%
\providecommand \bibitemStop [0]{}%
\providecommand \bibitemNoStop [0]{.\EOS\space}%
\providecommand \EOS [0]{\spacefactor3000\relax}%
\providecommand \BibitemShut  [1]{\csname bibitem#1\endcsname}%
\let\auto@bib@innerbib\@empty
\bibitem [{\citenamefont {de~Forcrand}\ and\ \citenamefont
  {Philipsen}(2007)}]{deForcrand:2006pv}%
  \BibitemOpen
  \bibfield  {author} {\bibinfo {author} {\bibfnamefont {P.}~\bibnamefont
  {de~Forcrand}}\ and\ \bibinfo {author} {\bibfnamefont {O.}~\bibnamefont
  {Philipsen}},\ }\href {\doibase 10.1088/1126-6708/2007/01/077} {\bibfield
  {journal} {\bibinfo  {journal} {JHEP}\ }\textbf {\bibinfo {volume} {0701}},\
  \bibinfo {pages} {077} (\bibinfo {year} {2007})},\ \Eprint
  {http://arxiv.org/abs/hep-lat/0607017} {arXiv:hep-lat/0607017 [hep-lat]}
  \BibitemShut {NoStop}%
\bibitem [{\citenamefont {Fromm}\ \emph {et~al.}(2012)\citenamefont {Fromm},
  \citenamefont {Langelage}, \citenamefont {Lottini},\ and\ \citenamefont
  {Philipsen}}]{Fromm:2011qi}%
  \BibitemOpen
  \bibfield  {author} {\bibinfo {author} {\bibfnamefont {M.}~\bibnamefont
  {Fromm}}, \bibinfo {author} {\bibfnamefont {J.}~\bibnamefont {Langelage}},
  \bibinfo {author} {\bibfnamefont {S.}~\bibnamefont {Lottini}}, \ and\
  \bibinfo {author} {\bibfnamefont {O.}~\bibnamefont {Philipsen}},\ }\href
  {\doibase 10.1007/JHEP01(2012)042} {\bibfield  {journal} {\bibinfo  {journal}
  {JHEP}\ }\textbf {\bibinfo {volume} {1201}},\ \bibinfo {pages} {042}
  (\bibinfo {year} {2012})},\ \Eprint {http://arxiv.org/abs/1111.4953}
  {arXiv:1111.4953 [hep-lat]} \BibitemShut {NoStop}%
\bibitem [{\citenamefont {Saito}\ \emph {et~al.}(2011)\citenamefont {Saito},
  \citenamefont {Ejiri}, \citenamefont {Aoki}, \citenamefont {Hatsuda},
  \citenamefont {Kanaya}, \citenamefont {Maezawa}, \citenamefont {Ohno},\ and\
  \citenamefont {Umeda}}]{Saito:2011fs}%
  \BibitemOpen
  \bibfield  {author} {\bibinfo {author} {\bibfnamefont {H.}~\bibnamefont
  {Saito}}, \bibinfo {author} {\bibfnamefont {S.}~\bibnamefont {Ejiri}},
  \bibinfo {author} {\bibfnamefont {S.}~\bibnamefont {Aoki}}, \bibinfo {author}
  {\bibfnamefont {T.}~\bibnamefont {Hatsuda}}, \bibinfo {author} {\bibfnamefont
  {K.}~\bibnamefont {Kanaya}}, \bibinfo {author} {\bibfnamefont
  {Y.}~\bibnamefont {Maezawa}}, \bibinfo {author} {\bibfnamefont
  {H.}~\bibnamefont {Ohno}}, \ and\ \bibinfo {author} {\bibfnamefont
  {T.}~\bibnamefont {Umeda}} (\bibinfo {collaboration} {WHOT-QCD}),\ }\href
  {\doibase 10.1103/PhysRevD.85.079902, 10.1103/PhysRevD.84.054502} {\bibfield
  {journal} {\bibinfo  {journal} {Phys. Rev.}\ }\textbf {\bibinfo {volume}
  {D84}},\ \bibinfo {pages} {054502} (\bibinfo {year} {2011})},\ \bibinfo
  {note} {[Erratum: Phys. Rev.D85,079902(2012)]},\ \Eprint
  {http://arxiv.org/abs/1106.0974} {arXiv:1106.0974 [hep-lat]} \BibitemShut
  {NoStop}%
\bibitem [{\citenamefont {Pisarski}\ and\ \citenamefont
  {Wilczek}(1984)}]{Pisarski:1983ms}%
  \BibitemOpen
  \bibfield  {author} {\bibinfo {author} {\bibfnamefont {R.~D.}\ \bibnamefont
  {Pisarski}}\ and\ \bibinfo {author} {\bibfnamefont {F.}~\bibnamefont
  {Wilczek}},\ }\href {\doibase 10.1103/PhysRevD.29.338} {\bibfield  {journal}
  {\bibinfo  {journal} {Phys. Rev.}\ }\textbf {\bibinfo {volume} {D29}},\
  \bibinfo {pages} {338} (\bibinfo {year} {1984})}\BibitemShut {NoStop}%
\bibitem [{\citenamefont {Butti}\ \emph {et~al.}(2003)\citenamefont {Butti},
  \citenamefont {Pelissetto},\ and\ \citenamefont {Vicari}}]{Butti:2003nu}%
  \BibitemOpen
  \bibfield  {author} {\bibinfo {author} {\bibfnamefont {A.}~\bibnamefont
  {Butti}}, \bibinfo {author} {\bibfnamefont {A.}~\bibnamefont {Pelissetto}}, \
  and\ \bibinfo {author} {\bibfnamefont {E.}~\bibnamefont {Vicari}},\ }\href
  {\doibase 10.1088/1126-6708/2003/08/029} {\bibfield  {journal} {\bibinfo
  {journal} {JHEP}\ }\textbf {\bibinfo {volume} {08}},\ \bibinfo {pages} {029}
  (\bibinfo {year} {2003})},\ \Eprint {http://arxiv.org/abs/hep-ph/0307036}
  {arXiv:hep-ph/0307036 [hep-ph]} \BibitemShut {NoStop}%
\bibitem [{\citenamefont {Meyer}(2015)}]{Meyer:2015}%
  \BibitemOpen
  \bibfield  {author} {\bibinfo {author} {\bibfnamefont {H.}~\bibnamefont
  {Meyer}},\ }\href@noop {} {\bibfield  {journal} {\bibinfo  {journal} {PoS}\
  }\textbf {\bibinfo {volume} {LAT2015}},\ \bibinfo {pages} {354} (\bibinfo
  {year} {2015})}\BibitemShut {NoStop}%
\bibitem [{\citenamefont {Rajagopal}\ and\ \citenamefont
  {Wilczek}(2000)}]{Rajagopal:2000wf}%
  \BibitemOpen
  \bibfield  {author} {\bibinfo {author} {\bibfnamefont {K.}~\bibnamefont
  {Rajagopal}}\ and\ \bibinfo {author} {\bibfnamefont {F.}~\bibnamefont
  {Wilczek}},\ }\href@noop {} {\  (\bibinfo {year} {2000})},\ \Eprint
  {http://arxiv.org/abs/hep-ph/0011333} {arXiv:hep-ph/0011333 [hep-ph]}
  \BibitemShut {NoStop}%
\bibitem [{\citenamefont {Bonati}\ \emph {et~al.}(2014)\citenamefont {Bonati},
  \citenamefont {de~Forcrand}, \citenamefont {D'Elia}, \citenamefont
  {Philipsen},\ and\ \citenamefont {Sanfilippo}}]{Bonati:2014kpa}%
  \BibitemOpen
  \bibfield  {author} {\bibinfo {author} {\bibfnamefont {C.}~\bibnamefont
  {Bonati}}, \bibinfo {author} {\bibfnamefont {P.}~\bibnamefont {de~Forcrand}},
  \bibinfo {author} {\bibfnamefont {M.}~\bibnamefont {D'Elia}}, \bibinfo
  {author} {\bibfnamefont {O.}~\bibnamefont {Philipsen}}, \ and\ \bibinfo
  {author} {\bibfnamefont {F.}~\bibnamefont {Sanfilippo}},\ }\href {\doibase
  10.1103/PhysRevD.90.074030} {\bibfield  {journal} {\bibinfo  {journal} {Phys.
  Rev.}\ }\textbf {\bibinfo {volume} {D90}},\ \bibinfo {pages} {074030}
  (\bibinfo {year} {2014})},\ \Eprint {http://arxiv.org/abs/1408.5086}
  {arXiv:1408.5086 [hep-lat]} \BibitemShut {NoStop}%
\bibitem [{\citenamefont {Ejiri}\ \emph {et~al.}(2016)\citenamefont {Ejiri},
  \citenamefont {Iwami},\ and\ \citenamefont {Yamada}}]{Ejiri:2015vip}%
  \BibitemOpen
  \bibfield  {author} {\bibinfo {author} {\bibfnamefont {S.}~\bibnamefont
  {Ejiri}}, \bibinfo {author} {\bibfnamefont {R.}~\bibnamefont {Iwami}}, \ and\
  \bibinfo {author} {\bibfnamefont {N.}~\bibnamefont {Yamada}},\ }\href
  {\doibase 10.1103/PhysRevD.93.054506} {\bibfield  {journal} {\bibinfo
  {journal} {Phys. Rev.}\ }\textbf {\bibinfo {volume} {D93}},\ \bibinfo {pages}
  {054506} (\bibinfo {year} {2016})},\ \Eprint
  {http://arxiv.org/abs/1511.06126} {arXiv:1511.06126 [hep-lat]} \BibitemShut
  {NoStop}%
\bibitem [{\citenamefont {de~Forcrand}\ \emph {et~al.}(2007)\citenamefont
  {de~Forcrand}, \citenamefont {Kim},\ and\ \citenamefont
  {Philipsen}}]{deForcrand:2007rq}%
  \BibitemOpen
  \bibfield  {author} {\bibinfo {author} {\bibfnamefont {P.}~\bibnamefont
  {de~Forcrand}}, \bibinfo {author} {\bibfnamefont {S.}~\bibnamefont {Kim}}, \
  and\ \bibinfo {author} {\bibfnamefont {O.}~\bibnamefont {Philipsen}},\
  }\href@noop {} {\bibfield  {journal} {\bibinfo  {journal} {PoS}\ }\textbf
  {\bibinfo {volume} {LAT2007}},\ \bibinfo {pages} {178} (\bibinfo {year}
  {2007})},\ \Eprint {http://arxiv.org/abs/0711.0262} {arXiv:0711.0262
  [hep-lat]} \BibitemShut {NoStop}%
\bibitem [{\citenamefont {Endrodi}\ \emph {et~al.}(2007)\citenamefont
  {Endrodi}, \citenamefont {Fodor}, \citenamefont {Katz},\ and\ \citenamefont
  {Szabo}}]{Endrodi:2007gc}%
  \BibitemOpen
  \bibfield  {author} {\bibinfo {author} {\bibfnamefont {G.}~\bibnamefont
  {Endrodi}}, \bibinfo {author} {\bibfnamefont {Z.}~\bibnamefont {Fodor}},
  \bibinfo {author} {\bibfnamefont {S.~D.}\ \bibnamefont {Katz}}, \ and\
  \bibinfo {author} {\bibfnamefont {K.~K.}\ \bibnamefont {Szabo}},\ }\href@noop
  {} {\bibfield  {journal} {\bibinfo  {journal} {PoS}\ }\textbf {\bibinfo
  {volume} {LAT2007}},\ \bibinfo {pages} {182} (\bibinfo {year} {2007})},\
  \Eprint {http://arxiv.org/abs/0710.0998} {arXiv:0710.0998 [hep-lat]}
  \BibitemShut {NoStop}%
\bibitem [{\citenamefont {Ding}\ and\ \citenamefont
  {Hegde}(2015)}]{Ding:2015pmg}%
  \BibitemOpen
  \bibfield  {author} {\bibinfo {author} {\bibfnamefont {H.-T.}\ \bibnamefont
  {Ding}}\ and\ \bibinfo {author} {\bibfnamefont {P.}~\bibnamefont {Hegde}}
  (\bibinfo {collaboration} {Bielefeld-BNL-CCNU}),\ }\href
  {http://inspirehep.net/record/1402370/files/arXiv:1511.00553.pdf} {\ \textbf
  {\bibinfo {volume} {LAT2015}} (\bibinfo {year} {2015})},\ \Eprint
  {http://arxiv.org/abs/1511.00553} {arXiv:1511.00553 [hep-lat]} \BibitemShut
  {NoStop}%
\bibitem [{\citenamefont {Takeda}\ \emph {et~al.}(2014)\citenamefont {Takeda},
  \citenamefont {Jin}, \citenamefont {Kuramashi}, \citenamefont {Nakamura},\
  and\ \citenamefont {Ukawa}}]{Takeda:2014nta}%
  \BibitemOpen
  \bibfield  {author} {\bibinfo {author} {\bibfnamefont {S.}~\bibnamefont
  {Takeda}}, \bibinfo {author} {\bibfnamefont {X.-Y.}\ \bibnamefont {Jin}},
  \bibinfo {author} {\bibfnamefont {Y.}~\bibnamefont {Kuramashi}}, \bibinfo
  {author} {\bibfnamefont {Y.}~\bibnamefont {Nakamura}}, \ and\ \bibinfo
  {author} {\bibfnamefont {A.}~\bibnamefont {Ukawa}},\ }\href@noop {}
  {\bibfield  {journal} {\bibinfo  {journal} {PoS}\ }\textbf {\bibinfo {volume}
  {LAT2014}},\ \bibinfo {pages} {196} (\bibinfo {year} {2014})},\ \Eprint
  {http://arxiv.org/abs/1411.1148} {arXiv:1411.1148 [hep-lat]} \BibitemShut
  {NoStop}%
\bibitem [{\citenamefont {Philipsen}\ and\ \citenamefont
  {Pinke}(2014)}]{Philipsen:2014rpa}%
  \BibitemOpen
  \bibfield  {author} {\bibinfo {author} {\bibfnamefont {O.}~\bibnamefont
  {Philipsen}}\ and\ \bibinfo {author} {\bibfnamefont {C.}~\bibnamefont
  {Pinke}},\ }\href {\doibase 10.1103/PhysRevD.89.094504} {\bibfield  {journal}
  {\bibinfo  {journal} {Phys. Rev.}\ }\textbf {\bibinfo {volume} {D89}},\
  \bibinfo {pages} {094504} (\bibinfo {year} {2014})},\ \Eprint
  {http://arxiv.org/abs/1402.0838} {arXiv:1402.0838 [hep-lat]} \BibitemShut
  {NoStop}%
\bibitem [{\citenamefont {Cuteri}\ \emph {et~al.}(2015)\citenamefont {Cuteri},
  \citenamefont {Pinke}, \citenamefont {Sciarra}, \citenamefont {Czaban},\ and\
  \citenamefont {Philipsen}}]{Cuteri:2015qkq}%
  \BibitemOpen
  \bibfield  {author} {\bibinfo {author} {\bibfnamefont {F.}~\bibnamefont
  {Cuteri}}, \bibinfo {author} {\bibfnamefont {C.}~\bibnamefont {Pinke}},
  \bibinfo {author} {\bibfnamefont {A.}~\bibnamefont {Sciarra}}, \bibinfo
  {author} {\bibfnamefont {C.}~\bibnamefont {Czaban}}, \ and\ \bibinfo {author}
  {\bibfnamefont {O.}~\bibnamefont {Philipsen}},\ }\href@noop {} {\  (\bibinfo
  {year} {2015})},\ \Eprint {http://arxiv.org/abs/1512.07180} {arXiv:1512.07180
  [hep-lat]} \BibitemShut {NoStop}%
\bibitem [{\citenamefont {Roberge}\ and\ \citenamefont
  {Weiss}(1986)}]{Roberge:1986mm}%
  \BibitemOpen
  \bibfield  {author} {\bibinfo {author} {\bibfnamefont {A.}~\bibnamefont
  {Roberge}}\ and\ \bibinfo {author} {\bibfnamefont {N.}~\bibnamefont
  {Weiss}},\ }\href {\doibase 10.1016/0550-3213(86)90582-1} {\bibfield
  {journal} {\bibinfo  {journal} {Nucl.Phys.}\ }\textbf {\bibinfo {volume}
  {B275}},\ \bibinfo {pages} {734} (\bibinfo {year} {1986})}\BibitemShut
  {NoStop}%
\bibitem [{\citenamefont {de~Forcrand}\ and\ \citenamefont
  {Philipsen}(2002)}]{deForcrand:2002ci}%
  \BibitemOpen
  \bibfield  {author} {\bibinfo {author} {\bibfnamefont {P.}~\bibnamefont
  {de~Forcrand}}\ and\ \bibinfo {author} {\bibfnamefont {O.}~\bibnamefont
  {Philipsen}},\ }\href {\doibase 10.1016/S0550-3213(02)00626-0} {\bibfield
  {journal} {\bibinfo  {journal} {Nucl.Phys.}\ }\textbf {\bibinfo {volume}
  {B642}},\ \bibinfo {pages} {290} (\bibinfo {year} {2002})},\ \Eprint
  {http://arxiv.org/abs/hep-lat/0205016} {arXiv:hep-lat/0205016 [hep-lat]}
  \BibitemShut {NoStop}%
\bibitem [{\citenamefont {D'Elia}\ and\ \citenamefont
  {Lombardo}(2003)}]{D'Elia:2002gd}%
  \BibitemOpen
  \bibfield  {author} {\bibinfo {author} {\bibfnamefont {M.}~\bibnamefont
  {D'Elia}}\ and\ \bibinfo {author} {\bibfnamefont {M.-P.}\ \bibnamefont
  {Lombardo}},\ }\href {\doibase 10.1103/PhysRevD.67.014505} {\bibfield
  {journal} {\bibinfo  {journal} {Phys.Rev.}\ }\textbf {\bibinfo {volume}
  {D67}},\ \bibinfo {pages} {014505} (\bibinfo {year} {2003})},\ \Eprint
  {http://arxiv.org/abs/hep-lat/0209146} {arXiv:hep-lat/0209146 [hep-lat]}
  \BibitemShut {NoStop}%
\bibitem [{\citenamefont {de~Forcrand}\ and\ \citenamefont
  {Philipsen}(2010)}]{deForcrand:2010he}%
  \BibitemOpen
  \bibfield  {author} {\bibinfo {author} {\bibfnamefont {P.}~\bibnamefont
  {de~Forcrand}}\ and\ \bibinfo {author} {\bibfnamefont {O.}~\bibnamefont
  {Philipsen}},\ }\href {\doibase 10.1103/PhysRevLett.105.152001} {\bibfield
  {journal} {\bibinfo  {journal} {Phys.Rev.Lett.}\ }\textbf {\bibinfo {volume}
  {105}},\ \bibinfo {pages} {152001} (\bibinfo {year} {2010})},\ \Eprint
  {http://arxiv.org/abs/1004.3144} {arXiv:1004.3144 [hep-lat]} \BibitemShut
  {NoStop}%
\bibitem [{\citenamefont {de~Forcrand}\ and\ \citenamefont
  {Philipsen}(2003)}]{deForcrand:2003vyj}%
  \BibitemOpen
  \bibfield  {author} {\bibinfo {author} {\bibfnamefont {P.}~\bibnamefont
  {de~Forcrand}}\ and\ \bibinfo {author} {\bibfnamefont {O.}~\bibnamefont
  {Philipsen}},\ }\href {\doibase 10.1016/j.nuclphysb.2003.09.005} {\bibfield
  {journal} {\bibinfo  {journal} {Nucl. Phys.}\ }\textbf {\bibinfo {volume}
  {B673}},\ \bibinfo {pages} {170} (\bibinfo {year} {2003})},\ \Eprint
  {http://arxiv.org/abs/hep-lat/0307020} {arXiv:hep-lat/0307020 [hep-lat]}
  \BibitemShut {NoStop}%
\bibitem [{\citenamefont {de~Forcrand}\ and\ \citenamefont
  {Philipsen}(2008)}]{deForcrand:2008vr}%
  \BibitemOpen
  \bibfield  {author} {\bibinfo {author} {\bibfnamefont {P.}~\bibnamefont
  {de~Forcrand}}\ and\ \bibinfo {author} {\bibfnamefont {O.}~\bibnamefont
  {Philipsen}},\ }\href {\doibase 10.1088/1126-6708/2008/11/012} {\bibfield
  {journal} {\bibinfo  {journal} {JHEP}\ }\textbf {\bibinfo {volume} {0811}},\
  \bibinfo {pages} {012} (\bibinfo {year} {2008})},\ \Eprint
  {http://arxiv.org/abs/0808.1096} {arXiv:0808.1096 [hep-lat]} \BibitemShut
  {NoStop}%
\bibitem [{\citenamefont {D'Elia}\ and\ \citenamefont
  {Sanfilippo}(2009)}]{D'Elia:2009qz}%
  \BibitemOpen
  \bibfield  {author} {\bibinfo {author} {\bibfnamefont {M.}~\bibnamefont
  {D'Elia}}\ and\ \bibinfo {author} {\bibfnamefont {F.}~\bibnamefont
  {Sanfilippo}},\ }\href {\doibase 10.1103/PhysRevD.80.111501} {\bibfield
  {journal} {\bibinfo  {journal} {Phys. Rev.}\ }\textbf {\bibinfo {volume}
  {D80}},\ \bibinfo {pages} {111501} (\bibinfo {year} {2009})},\ \Eprint
  {http://arxiv.org/abs/0909.0254} {arXiv:0909.0254 [hep-lat]} \BibitemShut
  {NoStop}%
\bibitem [{\citenamefont {Bonati}\ \emph {et~al.}(2011)\citenamefont {Bonati},
  \citenamefont {Cossu}, \citenamefont {D'Elia},\ and\ \citenamefont
  {Sanfilippo}}]{Bonati:2010gi}%
  \BibitemOpen
  \bibfield  {author} {\bibinfo {author} {\bibfnamefont {C.}~\bibnamefont
  {Bonati}}, \bibinfo {author} {\bibfnamefont {G.}~\bibnamefont {Cossu}},
  \bibinfo {author} {\bibfnamefont {M.}~\bibnamefont {D'Elia}}, \ and\ \bibinfo
  {author} {\bibfnamefont {F.}~\bibnamefont {Sanfilippo}},\ }\href {\doibase
  10.1103/PhysRevD.83.054505} {\bibfield  {journal} {\bibinfo  {journal}
  {Phys.Rev.}\ }\textbf {\bibinfo {volume} {D83}},\ \bibinfo {pages} {054505}
  (\bibinfo {year} {2011})},\ \Eprint {http://arxiv.org/abs/1011.4515}
  {arXiv:1011.4515 [hep-lat]} \BibitemShut {NoStop}%
\bibitem [{\citenamefont {Bonati}\ \emph {et~al.}(2016)\citenamefont {Bonati},
  \citenamefont {D'Elia}, \citenamefont {Mariti}, \citenamefont {Mesiti},
  \citenamefont {Negro},\ and\ \citenamefont {Sanfilippo}}]{Bonati:2016pwz}%
  \BibitemOpen
  \bibfield  {author} {\bibinfo {author} {\bibfnamefont {C.}~\bibnamefont
  {Bonati}}, \bibinfo {author} {\bibfnamefont {M.}~\bibnamefont {D'Elia}},
  \bibinfo {author} {\bibfnamefont {M.}~\bibnamefont {Mariti}}, \bibinfo
  {author} {\bibfnamefont {M.}~\bibnamefont {Mesiti}}, \bibinfo {author}
  {\bibfnamefont {F.}~\bibnamefont {Negro}}, \ and\ \bibinfo {author}
  {\bibfnamefont {F.}~\bibnamefont {Sanfilippo}},\ }\href@noop {} {\  (\bibinfo
  {year} {2016})},\ \Eprint {http://arxiv.org/abs/1602.01426} {arXiv:1602.01426
  [hep-lat]} \BibitemShut {NoStop}%
\bibitem [{\citenamefont {Nagata}\ and\ \citenamefont
  {Nakamura}(2011)}]{Nagata:2011yf}%
  \BibitemOpen
  \bibfield  {author} {\bibinfo {author} {\bibfnamefont {K.}~\bibnamefont
  {Nagata}}\ and\ \bibinfo {author} {\bibfnamefont {A.}~\bibnamefont
  {Nakamura}},\ }\href {\doibase 10.1103/PhysRevD.83.114507} {\bibfield
  {journal} {\bibinfo  {journal} {Phys.Rev.}\ }\textbf {\bibinfo {volume}
  {D83}},\ \bibinfo {pages} {114507} (\bibinfo {year} {2011})},\ \Eprint
  {http://arxiv.org/abs/1104.2142} {arXiv:1104.2142 [hep-lat]} \BibitemShut
  {NoStop}%
\bibitem [{\citenamefont {Alexandru}\ and\ \citenamefont
  {Li}(2013)}]{Alexandru:2013uaa}%
  \BibitemOpen
  \bibfield  {author} {\bibinfo {author} {\bibfnamefont {A.}~\bibnamefont
  {Alexandru}}\ and\ \bibinfo {author} {\bibfnamefont {A.}~\bibnamefont {Li}},\
  }\href@noop {} {\bibfield  {journal} {\bibinfo  {journal} {PoS}\ }\textbf
  {\bibinfo {volume} {LAT2013}},\ \bibinfo {pages} {208} (\bibinfo {year}
  {2013})},\ \Eprint {http://arxiv.org/abs/1312.1201} {arXiv:1312.1201
  [hep-lat]} \BibitemShut {NoStop}%
\bibitem [{\citenamefont {Wu}\ and\ \citenamefont {Meng}(2013)}]{Wu:2013bfa}%
  \BibitemOpen
  \bibfield  {author} {\bibinfo {author} {\bibfnamefont {L.-K.}\ \bibnamefont
  {Wu}}\ and\ \bibinfo {author} {\bibfnamefont {X.-F.}\ \bibnamefont {Meng}},\
  }\href {\doibase 10.1103/PhysRevD.87.094508} {\bibfield  {journal} {\bibinfo
  {journal} {Phys.Rev.}\ }\textbf {\bibinfo {volume} {D87}},\ \bibinfo {pages}
  {094508} (\bibinfo {year} {2013})},\ \Eprint {http://arxiv.org/abs/1303.0336}
  {arXiv:1303.0336 [hep-lat]} \BibitemShut {NoStop}%
\bibitem [{\citenamefont {Bach}\ \emph {et~al.}()\citenamefont {Bach},
  \citenamefont {Pinke}, \citenamefont {Sciarra} \emph {et~al.}}]{CL2QCD}%
  \BibitemOpen
  \bibfield  {author} {\bibinfo {author} {\bibfnamefont {M.}~\bibnamefont
  {Bach}}, \bibinfo {author} {\bibfnamefont {C.}~\bibnamefont {Pinke}},
  \bibinfo {author} {\bibfnamefont {A.}~\bibnamefont {Sciarra}},  \emph
  {et~al.},\ }\href@noop {} {\enquote {\bibinfo {title} {{\clqcd}},}\ }\bibinfo
  {howpublished} {\mbox{\url{https://github.com/CL2QCD/cl2qcd}}}\BibitemShut
  {NoStop}%
\bibitem [{\citenamefont {{Khronos Working Group}}()}]{opencl}%
  \BibitemOpen
  \bibfield  {author} {\bibinfo {author} {\bibnamefont {{Khronos Working
  Group}}},\ }\href@noop {} {\enquote {\bibinfo {title} {The {OpenCL}
  {Specification}},}\ }\bibinfo {note}
  {Http://www.khronos.org/registry/cl/}\BibitemShut {NoStop}%
\bibitem [{\citenamefont {Bach}\ \emph {et~al.}(2013)\citenamefont {Bach},
  \citenamefont {Lindenstruth}, \citenamefont {Philipsen},\ and\ \citenamefont
  {Pinke}}]{Bach:2012iw}%
  \BibitemOpen
  \bibfield  {author} {\bibinfo {author} {\bibfnamefont {M.}~\bibnamefont
  {Bach}}, \bibinfo {author} {\bibfnamefont {V.}~\bibnamefont {Lindenstruth}},
  \bibinfo {author} {\bibfnamefont {O.}~\bibnamefont {Philipsen}}, \ and\
  \bibinfo {author} {\bibfnamefont {C.}~\bibnamefont {Pinke}},\ }\href
  {\doibase 10.1016/j.cpc.2013.03.020} {\bibfield  {journal} {\bibinfo
  {journal} {Comput.Phys.Commun.}\ }\textbf {\bibinfo {volume} {184}},\
  \bibinfo {pages} {2042} (\bibinfo {year} {2013})},\ \Eprint
  {http://arxiv.org/abs/1209.5942} {arXiv:1209.5942 [hep-lat]} \BibitemShut
  {NoStop}%
\bibitem [{\citenamefont {Philipsen}\ \emph {et~al.}(2014)\citenamefont
  {Philipsen}, \citenamefont {Pinke}, \citenamefont {Sciarra},\ and\
  \citenamefont {Bach}}]{Philipsen:2014mra}%
  \BibitemOpen
  \bibfield  {author} {\bibinfo {author} {\bibfnamefont {O.}~\bibnamefont
  {Philipsen}}, \bibinfo {author} {\bibfnamefont {C.}~\bibnamefont {Pinke}},
  \bibinfo {author} {\bibfnamefont {A.}~\bibnamefont {Sciarra}}, \ and\
  \bibinfo {author} {\bibfnamefont {M.}~\bibnamefont {Bach}},\ }\href@noop {}
  {\bibfield  {journal} {\bibinfo  {journal} {PoS}\ }\textbf {\bibinfo {volume}
  {LAT2014}},\ \bibinfo {pages} {038} (\bibinfo {year} {2014})},\ \Eprint
  {http://arxiv.org/abs/1411.5219} {arXiv:1411.5219 [hep-lat]} \BibitemShut
  {NoStop}%
\bibitem [{\citenamefont {Bach}\ \emph {et~al.}(2011)\citenamefont {Bach},
  \citenamefont {Kretz}, \citenamefont {Lindenstruth},\ and\ \citenamefont
  {Rohr}}]{Bach2011a}%
  \BibitemOpen
  \bibfield  {author} {\bibinfo {author} {\bibfnamefont {M.}~\bibnamefont
  {Bach}}, \bibinfo {author} {\bibfnamefont {M.}~\bibnamefont {Kretz}},
  \bibinfo {author} {\bibfnamefont {V.}~\bibnamefont {Lindenstruth}}, \ and\
  \bibinfo {author} {\bibfnamefont {D.}~\bibnamefont {Rohr}},\ }\href {\doibase
  10.1007/s00450-011-0161-5} {\bibfield  {journal} {\bibinfo  {journal}
  {Computer Science - Research and Development}\ ,\ \bibinfo {pages} {1}}
  (\bibinfo {year} {2011})}\BibitemShut {NoStop}%
\bibitem [{\citenamefont {Rohr}\ \emph {et~al.}(2015)\citenamefont {Rohr},
  \citenamefont {Bach}, \citenamefont {Neskovic}, \citenamefont {Lindenstruth},
  \citenamefont {Pinke},\ and\ \citenamefont {Philipsen}}]{L-CSC}%
  \BibitemOpen
  \bibfield  {author} {\bibinfo {author} {\bibfnamefont {D.}~\bibnamefont
  {Rohr}}, \bibinfo {author} {\bibfnamefont {M.}~\bibnamefont {Bach}}, \bibinfo
  {author} {\bibfnamefont {G.}~\bibnamefont {Neskovic}}, \bibinfo {author}
  {\bibfnamefont {V.}~\bibnamefont {Lindenstruth}}, \bibinfo {author}
  {\bibfnamefont {C.}~\bibnamefont {Pinke}}, \ and\ \bibinfo {author}
  {\bibfnamefont {O.}~\bibnamefont {Philipsen}},\ }in\ \href@noop {} {\emph
  {\bibinfo {booktitle} {High Performance Computing (LNCS)}}},\ Vol.\ \bibinfo
  {volume} {{9137}}\ (\bibinfo {year} {2015})\BibitemShut {NoStop}%
\bibitem [{Note1()}]{Note1}%
  \BibitemOpen
  \bibinfo {note} {See github.com/dhesse/py-uwerr.}\BibitemShut {Stop}%
\bibitem [{\citenamefont {Wolff}(2004)}]{Wolff:2003sm}%
  \BibitemOpen
  \bibfield  {author} {\bibinfo {author} {\bibfnamefont {U.}~\bibnamefont
  {Wolff}} (\bibinfo {collaboration} {ALPHA}),\ }\href {\doibase
  10.1016/S0010-4655(03)00467-3, 10.1016/j.cpc.2006.12.001} {\bibfield
  {journal} {\bibinfo  {journal} {Comput. Phys. Commun.}\ }\textbf {\bibinfo
  {volume} {156}},\ \bibinfo {pages} {143} (\bibinfo {year} {2004})},\ \bibinfo
  {note} {[Erratum: Comput. Phys. Commun.176,383(2007)]},\ \Eprint
  {http://arxiv.org/abs/hep-lat/0306017} {arXiv:hep-lat/0306017 [hep-lat]}
  \BibitemShut {NoStop}%
\bibitem [{\citenamefont {Ferrenberg}\ and\ \citenamefont
  {Swendsen}(1989)}]{Ferrenberg:1989ui}%
  \BibitemOpen
  \bibfield  {author} {\bibinfo {author} {\bibfnamefont {A.~M.}\ \bibnamefont
  {Ferrenberg}}\ and\ \bibinfo {author} {\bibfnamefont {R.~H.}\ \bibnamefont
  {Swendsen}},\ }\href {\doibase 10.1103/PhysRevLett.63.1195} {\bibfield
  {journal} {\bibinfo  {journal} {Phys.Rev.Lett.}\ }\textbf {\bibinfo {volume}
  {63}},\ \bibinfo {pages} {1195} (\bibinfo {year} {1989})}\BibitemShut
  {NoStop}%
\bibitem [{\citenamefont {Borsanyi}\ \emph {et~al.}(2012)\citenamefont
  {Borsanyi} \emph {et~al.}}]{Borsanyi:2012zs}%
  \BibitemOpen
  \bibfield  {author} {\bibinfo {author} {\bibfnamefont {S.}~\bibnamefont
  {Borsanyi}} \emph {et~al.},\ }\href {\doibase 10.1007/JHEP09(2012)010}
  {\bibfield  {journal} {\bibinfo  {journal} {JHEP}\ }\textbf {\bibinfo
  {volume} {09}},\ \bibinfo {pages} {010} (\bibinfo {year} {2012})},\ \Eprint
  {http://arxiv.org/abs/1203.4469} {arXiv:1203.4469 [hep-lat]} \BibitemShut
  {NoStop}%
\bibitem [{\citenamefont {Binder}(1981)}]{Binder:1981sa}%
  \BibitemOpen
  \bibfield  {author} {\bibinfo {author} {\bibfnamefont {K.}~\bibnamefont
  {Binder}},\ }\href {\doibase 10.1007/BF01293604} {\bibfield  {journal}
  {\bibinfo  {journal} {Z.Phys.}\ }\textbf {\bibinfo {volume} {B43}},\ \bibinfo
  {pages} {119} (\bibinfo {year} {1981})}\BibitemShut {NoStop}%
\bibitem [{\citenamefont {Depta}(2015)}]{thesisDepta}%
  \BibitemOpen
  \bibfield  {author} {\bibinfo {author} {\bibfnamefont {F.}~\bibnamefont
  {Depta}},\ }\href@noop {} {\bibfield  {journal} {\bibinfo  {journal}
  {(Bachelor Thesis)}\ } (\bibinfo {year} {2015})},\ \bibinfo {note}
  {\url{http://th.physik.uni-frankfurt.de/~philipsen/theses/depta_ba.pdf}}\BibitemShut
  {NoStop}%
\bibitem [{\citenamefont {Iwasaki}\ \emph {et~al.}(1996)\citenamefont
  {Iwasaki}, \citenamefont {Kanaya}, \citenamefont {Kaya}, \citenamefont
  {Sakai},\ and\ \citenamefont {Yoshie}}]{Iwasaki:1996zt}%
  \BibitemOpen
  \bibfield  {author} {\bibinfo {author} {\bibfnamefont {Y.}~\bibnamefont
  {Iwasaki}}, \bibinfo {author} {\bibfnamefont {K.}~\bibnamefont {Kanaya}},
  \bibinfo {author} {\bibfnamefont {S.}~\bibnamefont {Kaya}}, \bibinfo {author}
  {\bibfnamefont {S.}~\bibnamefont {Sakai}}, \ and\ \bibinfo {author}
  {\bibfnamefont {T.}~\bibnamefont {Yoshie}},\ }\href {\doibase
  10.1103/PhysRevD.54.7010} {\bibfield  {journal} {\bibinfo  {journal} {Phys.
  Rev.}\ }\textbf {\bibinfo {volume} {D54}},\ \bibinfo {pages} {7010} (\bibinfo
  {year} {1996})},\ \Eprint {http://arxiv.org/abs/hep-lat/9605030}
  {arXiv:hep-lat/9605030 [hep-lat]} \BibitemShut {NoStop}%
\bibitem [{\citenamefont {Burger}\ \emph {et~al.}(2013)\citenamefont {Burger},
  \citenamefont {Ilgenfritz}, \citenamefont {Kirchner}, \citenamefont
  {Lombardo}, \citenamefont {Müller-Preussker}, \citenamefont {Philipsen},
  \citenamefont {Urbach},\ and\ \citenamefont {Zeidlewicz}}]{Burger:2011zc}%
  \BibitemOpen
  \bibfield  {author} {\bibinfo {author} {\bibfnamefont {F.}~\bibnamefont
  {Burger}}, \bibinfo {author} {\bibfnamefont {E.-M.}\ \bibnamefont
  {Ilgenfritz}}, \bibinfo {author} {\bibfnamefont {M.}~\bibnamefont
  {Kirchner}}, \bibinfo {author} {\bibfnamefont {M.~P.}\ \bibnamefont
  {Lombardo}}, \bibinfo {author} {\bibfnamefont {M.}~\bibnamefont
  {Müller-Preussker}}, \bibinfo {author} {\bibfnamefont {O.}~\bibnamefont
  {Philipsen}}, \bibinfo {author} {\bibfnamefont {C.}~\bibnamefont {Urbach}}, \
  and\ \bibinfo {author} {\bibfnamefont {L.}~\bibnamefont {Zeidlewicz}}
  (\bibinfo {collaboration} {tmfT}),\ }\href {\doibase
  10.1103/PhysRevD.87.074508} {\bibfield  {journal} {\bibinfo  {journal} {Phys.
  Rev.}\ }\textbf {\bibinfo {volume} {D87}},\ \bibinfo {pages} {074508}
  (\bibinfo {year} {2013})},\ \Eprint {http://arxiv.org/abs/1102.4530}
  {arXiv:1102.4530 [hep-lat]} \BibitemShut {NoStop}%
\bibitem [{\citenamefont {Brandt}\ \emph {et~al.}(2014)\citenamefont {Brandt},
  \citenamefont {Francis}, \citenamefont {Meyer}, \citenamefont {Philipsen},\
  and\ \citenamefont {Wittig}}]{Brandt:2013mba}%
  \BibitemOpen
  \bibfield  {author} {\bibinfo {author} {\bibfnamefont {B.~B.}\ \bibnamefont
  {Brandt}}, \bibinfo {author} {\bibfnamefont {A.}~\bibnamefont {Francis}},
  \bibinfo {author} {\bibfnamefont {H.~B.}\ \bibnamefont {Meyer}}, \bibinfo
  {author} {\bibfnamefont {O.}~\bibnamefont {Philipsen}}, \ and\ \bibinfo
  {author} {\bibfnamefont {H.}~\bibnamefont {Wittig}},\ }\href@noop {}
  {\bibfield  {journal} {\bibinfo  {journal} {PoS}\ }\textbf {\bibinfo {volume}
  {LAT2013}},\ \bibinfo {pages} {162} (\bibinfo {year} {2014})},\ \Eprint
  {http://arxiv.org/abs/1310.8326} {arXiv:1310.8326 [hep-lat]} \BibitemShut
  {NoStop}%
\bibitem [{\citenamefont {Jin}\ \emph {et~al.}(2015)\citenamefont {Jin},
  \citenamefont {Kuramashi}, \citenamefont {Nakamura}, \citenamefont {Takeda},\
  and\ \citenamefont {Ukawa}}]{Jin:2014hea}%
  \BibitemOpen
  \bibfield  {author} {\bibinfo {author} {\bibfnamefont {X.-Y.}\ \bibnamefont
  {Jin}}, \bibinfo {author} {\bibfnamefont {Y.}~\bibnamefont {Kuramashi}},
  \bibinfo {author} {\bibfnamefont {Y.}~\bibnamefont {Nakamura}}, \bibinfo
  {author} {\bibfnamefont {S.}~\bibnamefont {Takeda}}, \ and\ \bibinfo {author}
  {\bibfnamefont {A.}~\bibnamefont {Ukawa}},\ }\href {\doibase
  10.1103/PhysRevD.91.014508} {\bibfield  {journal} {\bibinfo  {journal} {Phys.
  Rev.}\ }\textbf {\bibinfo {volume} {D91}},\ \bibinfo {pages} {014508}
  (\bibinfo {year} {2015})},\ \Eprint {http://arxiv.org/abs/1411.7461}
  {arXiv:1411.7461 [hep-lat]} \BibitemShut {NoStop}%
\end{thebibliography}%


\end{document}